\documentclass[article,twocolumn,preprintnumbers,superscriptaddress,nofootinbib]{revtex4}

\usepackage{enumerate}
\usepackage{mathrsfs,amsmath,amssymb}
\usepackage{natbib}
\usepackage{graphicx}
\usepackage{cancel}
\usepackage[normalem]{ulem}
\usepackage[pdftex,bookmarks,linktocpage,pdfpagelabels,plainpages=false,hyperfigures,linkcolor=blue,citecolor=blue,urlcolor=blue]{hyperref} 
\hypersetup{colorlinks=true}

\newcommand\be{\begin{equation}}
\newcommand\ee{\end{equation}}
\newcommand\bea{\begin{eqnarray}}
\newcommand\eea{\end{eqnarray}}

\newcommand\cevens{CE$\nu$NS }

\begin{document}
\bibliographystyle{apsrev4-1}

\title{Observing the Migdal effect from nuclear recoils of neutral particles with liquid xenon and argon detectors}

\author{Nicole F.~Bell}
\affiliation{ARC Centre of Excellence for Dark Matter Particle Physics, School of Physics, The University of Melbourne, Victoria 3010, Australia}

\author{James B.~Dent} 
\affiliation{Department of Physics, Sam Houston State University, Huntsville, TX 77341, USA}

\author{Rafael~F.~Lang}
\affiliation{Department of Physics and Astronomy, Purdue University, West Lafayette, IN 47907, USA}

\author{Jayden L.~Newstead}
\affiliation{ARC Centre of Excellence for Dark Matter Particle Physics, School of Physics, The University of Melbourne, Victoria 3010, Australia}
\affiliation{Department of Physics and Astronomy, Purdue University, West Lafayette, IN 47907, USA}

\author{Alexander C.~Ritter}
\affiliation{ARC Centre of Excellence for Dark Matter Particle Physics, School of Physics, The University of Melbourne, Victoria 3010, Australia}

\begin{abstract}
In recent years, dark matter direct detection experiments have spurred interest in the Migdal effect, where it is employed to extend their sensitivity to lower dark matter masses. Given the lack of observation of the Migdal effect, the calculation of the signal is subject to large theoretical uncertainties. It is therefore desirable to attempt a first measurement of the Migdal effect, and to test the theoretical predictions of the Migdal effect for the calibration of the experimental response to a potential dark matter signal. In this work, we explore the feasibility of observing the Migdal effect in xenon and argon. We carry out proof-of-concept calculations for low-energy neutrons from a filtered source, and using a reactor, the Spallation Neutron Source, or $^{51}$Cr as potential neutrino sources. We perform a detector simulation for the xenon target and find that, with available technology, the low-energy neutron source is the most promising, requiring only a modest neutron flux, detector size, and exposure period.
\end{abstract}

\maketitle

\section{Introduction}

The Migdal effect, first proposed by A.B.~Migdal over 80 years ago~\cite{Migdal:1941}, is the ionization of an atom following a nuclear recoil. In the frame of the electron cloud, the Coulomb potential of the nucleus is perturbed, leading to both excitation and ionization of the electron cloud. The Migdal effect occurs in isolated atomic systems and is not caused by any in-medium effects, though the Migdal effect will be altered due to interactions with neighboring atoms. Presently, observations of the Migdal effect have been limited to nuclear decay processes~\cite{PhysRevC.11.1740,PhysRevLett.108.243201}.

The Migdal effect has been explored for some time in the context of dark matter-nucleus scattering~\cite{Vergados:2004bm,Ejiri:2005aj,Moustakidis:2005gx,Bernabei:2007gr,Vergados:2013raa}. Recently, there has been increasing interest~\cite{Ibe:2017yqa,Dolan:2017xbu,Bell:2019egg,Baxter:2019pnz,Essig:2019xkx} in exploiting the Migdal effect to push experimental sensitivity to sub-GeV dark matter masses~\cite{Akerib:2018hck,Armengaud:2019kfj,Aprile:2019jmx,Wang:2019wwo,Bell:2021zkr}. The sensitivity is increased by two separate characteristics of the Migdal effect: the inelastic kinematics of the scattering and the nature of the energy deposition. Dark matter gravitationally bound to the Milky Way has a maximum speed of $\sim 750$ km/s in the lab frame~\cite{Necib:2021vxr}. This places an upper limit on the maximum energy transferred to the atom through an elastic nuclear recoil, which falls with the dark matter mass. However, including the effects from the inelastic Migdal recoil, the dark matter kinetic energy is more efficiently transferred into the kinetic and potential energy of the atomic electron cloud. Thus a Migdal recoil results in more energy being deposited into electronic energy, which can extend the recoil spectrum to higher energies than the nuclear recoil. This allows for the possibility of an observable signal even when the nuclear recoil is below the experimental threshold. Additionally, while a nuclear recoil signal is quenched, with only a fraction ($\sim$15\% in xenon) being detectable, electronic energy is detected with very high efficiency. The combination of these two effects has allowed xenon experiments to set world-leading bounds on the interactions of light dark matter despite their comparatively high energy thresholds~\cite{Aprile:2019jmx}.

Having not been observed due to nuclear scattering, the probability of the Migdal effect is subject to large theoretical uncertainties that are difficult to quantify (e.g.~\cite{Baxter:2019pnz}). Given the target-dependent nature of the Migdal effect, the interpretation of experimental results involving the Migdal effect from multiple targets is challenging. Therefore, it is important to measure and calibrate the Migdal effect directly via an independent method for each detector of interest. There are now experimental efforts underway to observe and calibrate the Migdal effect using neutron scattering~\cite{Majewski:2021}. Such a calibration is also needed for the unambiguous interpretation of low-energy neutrino scattering data~\cite{Liao:2021yog}. An observation could be achieved with either neutrons or neutrinos as the projectile. Both neutrons and neutrinos will dominantly interact with the nucleus and thus induce minimal electronic recoil backgrounds. While neutrinos do interact with electrons, for low incident neutrino energies the coherent elastic neutrino-nucleus (CE$\nu$NS) rate dominates over the electron scattering rate.

In this paper we provide a proof-of-concept calculation to assess the feasibility of observing the Migdal effect in detectors with liquid xenon and argon targets. We then provide detailed modeling of the detector response for a xenon time projection chamber (TPC) using the NEST simulation package. The paper is organized as follows: In section~\ref{sec:mig} we briefly review the calculation of the Migdal effect in the context of neutron and neutrino scattering. In sections~\ref{sec:neutron} and~\ref{sec:neutrino} we investigate the feasibility of observing the Migdal effect in xenon due to neutrons and neutrinos, respectively. In section~\ref{sec:detSim} we detail our detector model and perform NEST simulations to understand the detector response to Migdal events.

\begin{figure*}[!htbp]
\includegraphics[width=.4\textwidth]{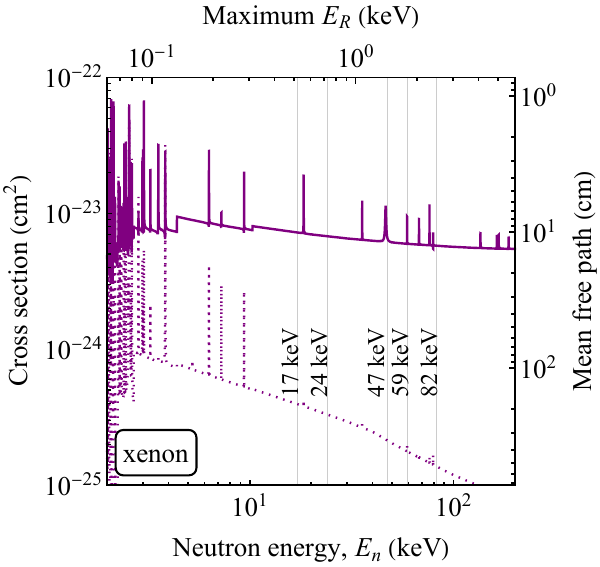}\hfill
\includegraphics[width=.4\textwidth]{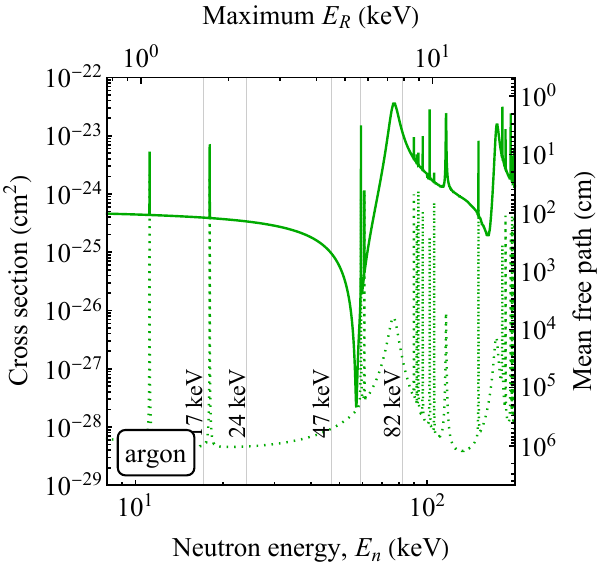}
\caption{The elastic (solid) and neutron capture (dotted) cross sections in the lab frame for neutrons scattering on xenon (left) and argon (right), as a function of the incoming neutron energy. For xenon, a weighted average was taken over the naturally-occurring isotopes. Data obtained from~\cite{JENDL4.0:2011} with line widths characteristic of 300~K.}
\label{fig:neutron_CS}
\end{figure*}

\section{The Migdal effect} \label{sec:mig}
In this section we review the calculation of the Migdal effect as it was presented in~\cite{Ibe:2017yqa}. The Migdal effect occurs as a consequence of the recoiling nucleus and so the differential rate of Migdal recoils is the product of the nuclear recoil rate for the 2-to-2 scattering processes $\nu+N\rightarrow \nu + N$ or $n+N\rightarrow n+N$, with the ionization rate, $Z_{\rm ion}$:
\bea
\label{eq:migdal_rate}
\frac{d^2R}{dE_{\rm NR}dE_i} &=& \frac{d^2R_{i T}}{dE_{\rm NR}dE_i} \times |Z_{{\rm{ion}}}|^2.
\eea
where $E_i$ is the incident particle energy and $E_{\rm NR}$ is the nuclear recoil energy. The ionization rate is given in terms of the ionization probability $p^c_{q_e}(n\ell\rightarrow(E_e))$
\bea
\label{eq:znl}
|Z_{{\rm{ion}}}|^2 = \frac{1}{2\pi}\sum_{n,\ell}\int dE_e\frac{d}{dE_e}p^c_{q_e}(n\ell\rightarrow(E_e)).
\eea
Atomic excitation is also possible due to nuclear recoils, however the ionization probability dominates over this effect in the energy region we consider here and so we neglect this effect. In~\cite{Ibe:2017yqa} the transition amplitudes are calculated in the single-electron approximation for atomic eigenstates boosted by a Galilean transformation, whose wavefunctions are computed using the Dirac-Hartree-Fock method. In our analysis we include the Migdal electrons originating from the $n=3,4,5$ ($n=1,2,3$) shells of xenon (argon). The detected energy spectrum is then obtained by summing the contributions from the nuclear recoil $E_R$, the ejected electron energy $E_e$, and the atomic de-excitation energy $E_{nl}$:
\be
E_{\mathrm{det}} = \mathcal{L} E_R + E_e + E_{nl}
\ee
where we have included a nuclear recoil quenching factor, $\mathcal{L}$. The differential event rate in terms of observed energy is obtained by integrating Eq.~(\ref{eq:migdal_rate}) over the atomic recoil energies and enforcing energy conservation. 

\begin{figure*}[t!]
\includegraphics[width=.4\textwidth]{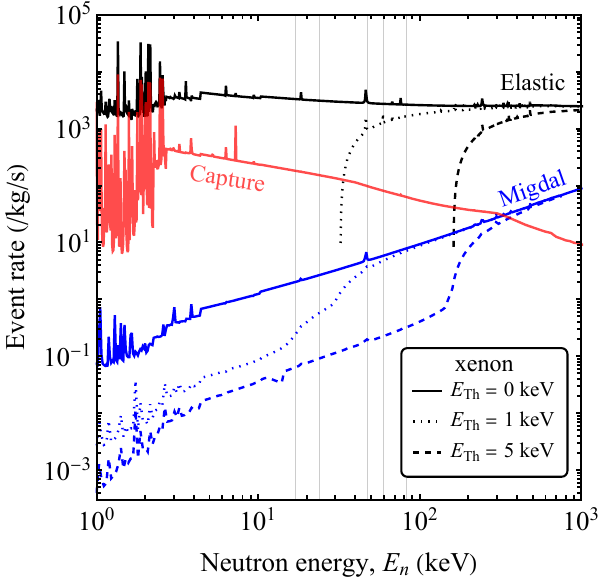}\hfill
\includegraphics[width=.4\textwidth]{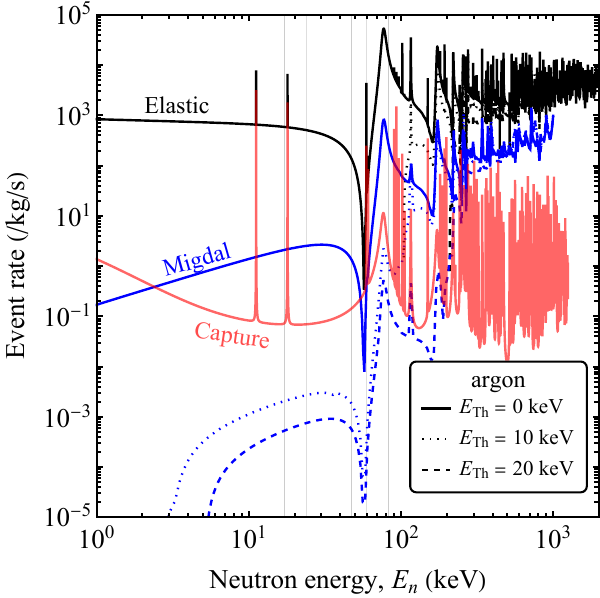}
\caption{The total elastic (black), Migdal (blue) and neutron capture (red) rates in xenon (left) and argon (right), as a function of the incoming neutron energy with a flux of 100 neutrons/cm$^2$/s (with the spectrum assumed to be a $\delta$ function at each energy). Line widths are characteristic of 300~K. The rates are integrated above three different benchmark detector thresholds: none, low (Xe: 1~keV$_{\text{NR}}$, Ar: 10~keV$_{\text{NR}}$) and high (Xe: 5~keV$_{\text{NR}}$, Ar: 20~keV$_{\text{NR}}$), shown in solid, dotted and dashed respectively. The gray vertical lines correspond to potential neutron beam energies of (17, 24, 47, 59, 82)~keV, from left to right.}
\label{fig:neutron_rates}
\end{figure*}

\section{Neutron scattering} 
\label{sec:neutron}

In this section we detail the calculation of the Migdal effect from elastic scattering of neutrons from xenon and argon nuclei as well as the intrinsic backgrounds. This has been previously explored in~\cite{Nakamura:2020kex}, where it was proposed to use gaseous detectors and neutrons of 565~keV energy. Our approach is different in two important ways. Firstly, we will consider lower-energy neutrons, which more closely resembles the kinematics of low-mass dark matter scattering. Secondly, we use liquid noble detectors, therefore measuring the Migdal response for the atomic system in liquid, as this is the target situation used in the most sensitive dark matter detectors. Following~\cite{PhysRevA.73.032722} we assume that the impulse (or sudden) approximation is valid and that we can ignore electromagnetic interactions of the neutron with the electron cloud.

\subsection{Elastic scattering}
Neutron-nucleus scattering is mediated by very short range meson-exchange currents which can be treated as contact interactions for the momentum transfers considered here. This simple picture is complicated by neutron capture processes that induce resonances in the elastic cross section. Detailed calculations of these cross sections have being carried out using the POD code and we make use of the results published in the JENDL-4.0 library~\cite{JENDL4.0:2011}. This library provides a good fit to the calibration data obtained by the LUX collaboration~\cite{Akerib:2016mzi}. We take the total elastic scattering cross section as the average of the cross section for all xenon isotopes (weighted by their naturally occurring abundances). This cross section is shown in Fig.~\ref{fig:neutron_CS} across a wide range of incoming neutron energies. The cross section exhibits many sharp peaks where the elastic amplitude interferes with the inelastic amplitude. This is available for a temperature of 300~K, which makes lines slightly wider than in the actual cryogenic liquid, a small systematic that does not impact our results, or if anything, renders them conservative. Note that the elastic cross section for argon was obtained via subtraction of the inelastic cross section from the total cross section and this can sometimes produce artificial structure in the elastic cross section. The JENDL-4.0 library also provides the angular distribution of scattered neutrons in the center of momentum frame, $f(\cos{\theta_{\mathrm{CM}}})$, which we use to infer the differential cross section in the lab frame with respect to the recoil energy, $E_R$:
\be
\frac{dR}{dE_R} = \sigma_{\mathrm{elastic}}(E_n) f(\cos{\theta_{\mathrm{CM}}}) \frac{d \cos{\theta}}{dE_R}.
\ee

Choosing an appropriate incoming neutron energy involves balancing several requirements. Ideally, the neutron energy would produce low energy nuclear recoils, be away from resonances, and can be produced as a mono-energetic neutron beam. Low energy nuclear recoils are desired so that most elastic scatters are below the detector threshold, enhancing our ability to identify Migdal events. However, low energy recoils have a smaller probability of causing a Migdal event. The kinematics of the scattering can be approximated via the formula~\cite{Akerib:2016mzi},
\be
\label{eq:ENRneutron}
E_{\rm NR} = E_n \frac{2 m_n m_T (1-\cos{\theta_{\rm~cm})}}{(m_n+m_T)^2}.
\ee
\be
\label{eq:ENRneutron}
E_{\rm R,max} =  \frac{4 \mu_T^2 E_n }{m_n m_T}
\ee
where $m_n$ and $m_T$ are the neutron and target nuclei masses, respectively, and $\mu_{\rm T}$ is the reduced mass of the neutron/nuclear target system $m_nm_{\rm T}/(m_n + m_{\rm T})$.
For liquid xenon detectors, nuclear recoil thresholds of $\mathcal{O}(1)$~keV have been achieved with small acceptance probability, which climbs to around 50\% at $\sim5$~keV~\cite{LUX:2016ggv,XENON:2018voc}. Argon detectors can also be sensitive to $\mathcal{O}(1)$~keV recoils, however pulse-shape cuts performed to remove electronic backgrounds can raise this to 10~keV~\cite{COHERENT:2020iec}, or even $\sim50$~keV~\cite{DarkSide:2018kuk}. Therefore, to keep the majority of elastic neutron scatters below threshold ($\sim1$~keV), neutron energies below 30~keV and 100~keV are required in xenon and argon, respectively. Neutrons of this energy are also below the inelastic scattering threshold and thus won't excite the nucleus - another potential background to the Migdal calibration. The elastic neutron-xenon cross section between the energies of $\sim 10-3000$~eV is dominated by resonances, while the region around $\sim10-200~\mathrm{keV}$ is relatively sparse. Except for a few major features, this region is also relatively clean in argon, while in the range of 100~keV-50~MeV, the cross section exhibits a lot of structure. For neutrons in this energy range, their velocity is around 1000 times the velocity of xenon atoms at 175~K (a typical temperature of noble liquids in a TPC). Therefore we can safely ignore the effect of temperature on the cross section and resonances.

Low energy mono-energetic neutron beams can be produced from reactor neutrons using filters~\cite{Barbeau:2007qh} or from proton beams using threshold nuclear reactions and filters~\cite{Joshi:2014oda}. A selection of the energies these sources can produce have been illustrated in Fig.~\ref{fig:neutron_CS}. The reactor source is able to deliver $\sim 10^8$ neutrons/hour from a megawatt reactor while the beam source can deliver $\sim 10^5$ neutrons/hour for proton beam currents of $600$~nA. In both cases fluxes could be increased by a factor of 10 without much trouble. The demonstrated beam size for reactor neutrons is 5.9~cm (full-width at half max) while the beam source was restricted to a 2~cm square path (no detailed beam profile data was obtained in~\cite{Joshi:2014oda}). A smaller beam could enable tighter fiducialization of the detector which would allow for more self-shielding to reduce external electronic recoil backgrounds.

Assuming an average flux of 100~neutrons/cm$^2$/s, the total event rate for Migdal events and nuclear recoils is given as a function of the neutron energy in Fig.~\ref{fig:neutron_rates} (see appendix~\ref{app:rates} for an example differential rate). At this flux, the Migdal effect could be feasibly observed across a wide range of energies. These graphs exhibit the increasing probability of Migdal events as the incident neutron energy is increased. They also highlight that once the elastic nuclear recoil energy is above detector threshold it will dominate the Migdal rate by orders of magnitude. This necessitates the need to ascertain how well a detector can distinguish a Migdal event from a nuclear recoil. This calculation also shows that the ratio of Migdal to nuclear recoil events is higher in argon than in xenon. However, the higher recoil threshold of argon detectors greatly suppresses the rate.

For our purposes we will assume an incoming neutron energy of 17~keV, giving a maximum nuclear recoil of 0.5~keV.  With such 17~keV neutrons the cross section of elastic scattering on xenon is 7.2~barn, giving a mean free path in liquid xenon of $\sim 10$~cm. Therefore, an $\mathcal{O}(1)$ fraction of incoming neutrons will elastically scatter within the detector volume. In argon, the mean free path is 120~cm, so a smaller fraction of the neutrons would scatter within the same volume. This, however, could be of benefit as it would reduce multiple scattering of neutrons within the detector. 

\subsection{Neutron capture}
Dark matter searches using TPCs have shown that internal backgrounds from detector materials and target contamination can be limited to around 100 events/tonne/yr/keV below 25 keV~\cite{XENON:2020rca}. Given the potentially large rate of Migdal events we will assume this background to be subdominant compared to the intrinsic and external backgrounds. The consideration of external backgrounds due to cosmic rays and environmental radioactivity is beyond the scope of this work, however they can be mitigated via the same methods we suggest later in this section. The intrinsic background to a neutron-beam Migdal calibration is radiative neutron capture and the subsequent $\beta$ and electron capture decays. The total neutron capture rates are shown in Fig.~\ref{fig:neutron_rates}. This rate neglects multiple scattering of neutrons within the detector and so should be regarded as a lower limit. For xenon, at neutron energies of 10-20~keV, the isotope-averaged neutron capture rate is only an order of magnitude smaller than the elastic scattering rate. On the other hand, argon enjoys a capture rate four orders of magnitude smaller than the elastic rate - below even the Migdal rate (if the sharp resonances in this energy range can be avoided). Given the relatively small rate of neutron capture in argon, we expect its contribution to the electronic recoil background to be negligible. We therefore focus the rest of this discussion on xenon. 

In xenon, the leading contribution to radiative neutron capture is due to $^{129}$Xe, followed by $^{131}$Xe. The $\gamma$-ray emission is mostly prompt with energies in the hundreds of~keV to~MeV range (see appendix~\ref{app:backgrounds} for details). The lower-energy $\gamma$-rays will have short mean free paths, below the detector resolution of a few millimeters, and will be dominantly photo-absorbed. $\gamma$-rays above a few hundred~keV can travel 1-10~cm and will dominantly Compton scatter, contributing to the low-energy electronic recoil background. This background can be reduced through cuts on energy and multiple scatters.

Neutron capture produces the unstable isotopes $^{125}$Xe, $^{127}$Xe, $^{133}$Xe, $^{135}$Xe and $^{137}$Xe, with half lives ranging from minutes to~days. The first two decay via electron capture while the other three $\beta$ decay. Typical $\beta$ decay energies are $>$MeV, which can be vetoed based on energy (see appendix~\ref{app:backgrounds} for details). On the other hand, electron capture induces Auger decay and subsequent $\gamma$ decays of iodine in the 10-100~keV range. The rates of these backgrounds initially grows linearly with the abundance of the unstable isotopes, eventually (after hundreds of hours) reaching a steady-state when the decay rate equals the production rate.  This slow rise time of these backgrounds allows for three potential mitigation techniques. A pulsed neutron beam source would allow one to trigger the Migdal signal on each beam pulse. The source could also be cycled on and off on the time scale of~days, allowing the unstable isotopes to decay away (at a cost of live-time). Alternately, a large amount of xenon could be cycled through the active volume of the detector, effectively diluting the abundance of unstable isotopes.

\section{Neutrino scattering} \label{sec:neutrino}
The Migdal effect due to \cevens ~has previously been treated in~\cite{Ibe:2017yqa} and~\cite{Bell:2019egg}. These previous works considered solar neutrinos as a source, finding Migdal rates below 10~events/tonne/year/keV, too small to be measured and distinguished from the neutrino-electron scattering rate. In this section we explore the feasibility of observing the Migdal effect from three different neutrino sources, not previously considered: nuclear reactors, the Spallation Neutron Source (SNS) and a radioisotope source chromium-51. We will assume optimistic characteristics for these neutrino sources to assess whether such a measurement is potentially viable.

\cevens is a neutral-current process with differential cross section:
\be
\frac{d\sigma}{d E_R} = \frac{G_F^2 m_T }{\pi}Q_w^2\left(1-\frac{m_T E_R}{2E_\nu^2}\right)F^2(E_R),
\ee
where $G_F$ is Fermi constant, $m_T$ and $Q_w$ are the mass and weak charge of the target nuclei, $E_R$ is the nuclear recoil energy and $E_\nu$ is the incoming neutrino energy. The form factor, $F(E_R)$, accounts for the loss of coherence at larger momentum transfers which we take to be of the Helm form~\cite{Helm:1956zz}. The coherent nature of the interaction implies a scaling of the cross section with the number of nucleons squared, but the relatively small Weak charge of the proton means that the scaling is closer to the number of neutrons squared. This implies that large atomic mass targets are favored for their neutron-rich nuclei. Here we will consider two targets: xenon, which benefits from being neutron rich, and argon, for which higher recoil energies are possible due to its lighter mass.

\begin{table}[b]
    \centering
    \begin{tabular}{|c|c|c|c|}
    \hline
        source &  flux (/cm$^2$/s)  & max $E_\nu$ (MeV) & max $E_R^{\text{Xe}}$ (keV) \\
        \hline 
        nuclear reactor & $1.5\times10^{13}$  &  10     & 1.7\\
        SNS             & $4.2\times10^6$     &  52.8   & 47\\
        $^{51}$Cr       & $4.8\times10^{13}$  & 0.746   & 0.01\\
         \hline
    \end{tabular}
    \caption{Characteristics of the neutrino sources considered in this work}
    \label{tab:nuSources}
\end{table}

The differential event rate per unit detector mass can be calculated from
\be
\frac{d^2 R}{d E_\nu\, d E_R} = \frac{1}{m_T}\frac{d\sigma}{d E_R}\frac{d\phi_{\nu,i}}{d E_\nu}\, \Theta\left(E_{R,\mathrm{max}}(E_\nu)-E_R\right)
\label{eq:diffRate}
\ee
where $\phi_{\nu,i}$ is the flux of the $i$th neutrino species, and $\Theta$ is the Heaviside step function which restricts $E_R$ to be less than the maximum value, corresponding to back-to-back scattering:
\be
E_{R,\mathrm{max}} =  \frac{2 E_\nu^2}{m_T+2E_\nu^2}.
\ee

\begin{figure}[t]
\includegraphics[angle=0,width=.4\textwidth]{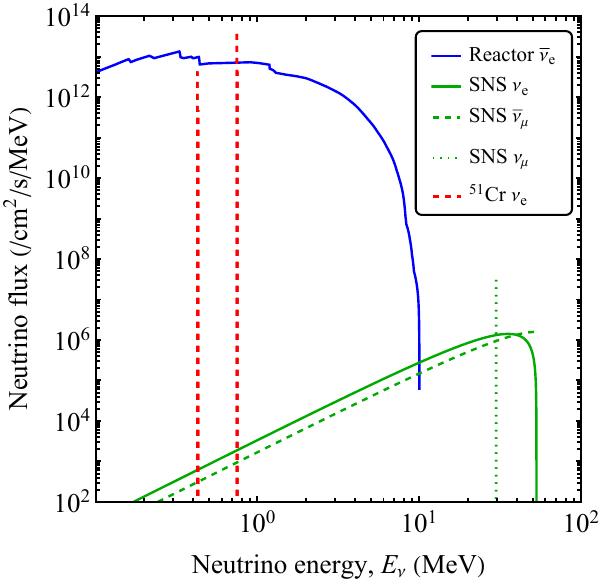}
\caption{The spectra of neutrino energies for the three neutrino sources considered in this work}
\label{fig:nuFluxes}
\end{figure}
\begin{figure*}[t!]
\includegraphics[angle=0,width=.4\textwidth]{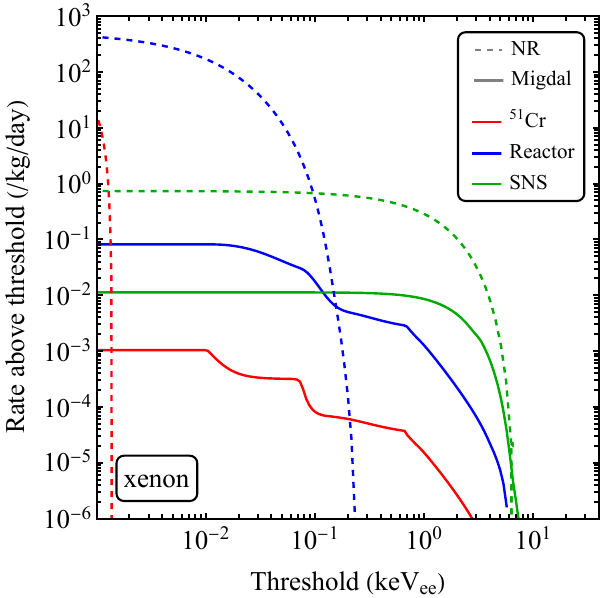}\hfill
\includegraphics[angle=0,width=.4\textwidth]{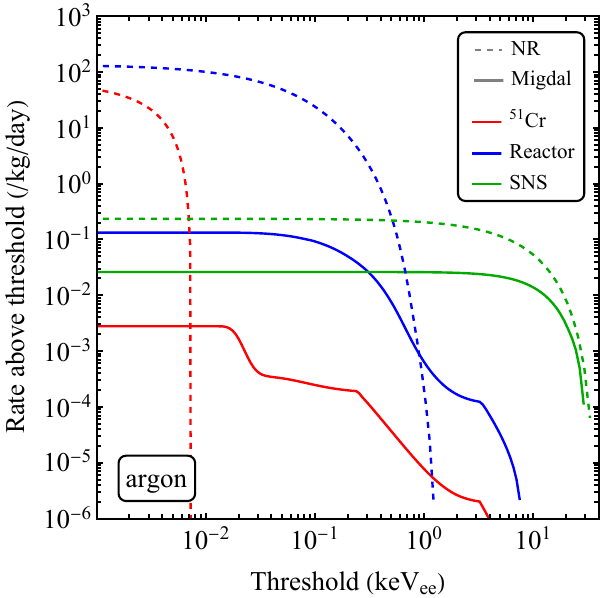}\\
\caption{The integrated rate of neutrino induced nuclear recoils (dashed) and Migdal events (solid) in xenon (left) and argon (right) from three neutrino sources: nuclear reactors, the SNS and chromium-51.}
\label{fig:NuRates}
\end{figure*}
Nuclear reactors contain fission products which undergo $\beta$ decay, causing them to emit a large flux of low-energy anti-neutrinos ($< 10$~MeV). Reactor sources are currently the subject of ongoing \cevens  experiments. Here we will consider a 1~GW reactor at a distance of 10~m.
The SNS produces spallation neutrons via a proton beam impinging on a mercury target. This process also produces a large number of pions that are promptly slowed down in the dense target. These pions then decay at rest (DAR), producing muons which decay in flight. This chain produces three neutrinos with energies below $m_\mu$/2. The COHERENT collaboration used the SNS to demonstrate the first ever observation of CE$\nu$NS. To estimate the flux of neutrinos from the SNS we assume a 1.4~MW proton beam at 0.984~GeV per proton and use a pion yield of 8.5\%~\cite{Akimov:2021geg}. We assume a detector distance of 12~m.
Some nuclear decays produce monoenergetic neutrinos, providing a useful calibration source. The GALLEX experiment used a 62~PBq sample of chromium-51~\cite{ANSELMANN1995440}, which decays via electron capture with a half-life of 27.7~days. The decay proceeds to either the 7/2$^-$ ground state or the 5/2$^{-}$ excited state of the $^{51}$V nucleus, producing mono-energetic neutrino lines of 745.8~keV, 750.7~keV, 425.7~keV and 430.6~keV, with branching fractions of 81\%, 9\%, 9\% and 1\% respectively~\cite{CRIBIER1988574}. To assess the plausibility of such a calibration source we will optimistically assume an activity of 60~PBq and a source-detector distance of 1~m.
A summary of the source characteristics are given in table~\ref{tab:nuSources} and the spectra of neutrinos they produce are given in Fig.~\ref{fig:nuFluxes}. We calculated the expected \cevens and Migdal rates for each of the neutrino sources and for detectors based on both xenon and argon. The rates are displayed as the integrated rate above a given threshold in Fig.~\ref{fig:NuRates} (for the differential rates see appendix~\ref{app:rates}). These results show that even with the optimistic source characteristics assumed here, observing Migdal events from neutrinos would require exposures of order 1-10~tonne-years. As with the neutron scattering case, the ratio of Migdal to nuclear recoil events is higher in argon, leading to Migdal rates that exceed those in xenon even though the nuclear recoil rate is smaller in argon.

\section{Detector simulation}
\label{sec:detSim}

In this section we present a rough design of a detector that is capable of observing the Migdal effect. We focus on a xenon target because the absolute Migdal rate is higher in xenon due to higher nuclear recoil cross sections and lower energy thresholds. The smaller mean free path of the neutrons also enables a more compact detector design. We consider this analysis exploratory, with the goal of evaluating whether such a detector can feasibly observe a sufficient number of Migdal events and whether they can be distinguished from the irreducible nuclear recoil background. As such we leave a statistical analysis that would include an estimation of electronic recoil background rates from external sources (e.g. Compton scattering of gamma-rays from the nuclear reactor) and intrinsic backgrounds (e.g. radiative neutron capture) to a future work.

The most sensitive dark matter detectors built are liquid xenon TPCs. Such detectors are sensitive to $\mathcal{O}$(keV) nuclear recoils and can be scaled to multi-tonne target masses. TPCs operate in a dual-phase configuration, typically with a cylindrical shape, where a drift field is applied to the liquid phase and a stronger extraction field applied to pull charges into the gas phase. The larger liquid phase provides the main active detector medium, with a smaller gas phase above it. Photomultiplier tubes (PMTs) are placed in an array above and below the detector volume. When a nuclear or electronic recoil occurs within the liquid phase, quanta of photons and ions are produced, with the total number of quanta being proportional to the recoil energy. The photons are emitted as prompt scintillation light, at a wavelength that xenon is relatively transparent to. This allows the light to propagate out of the xenon and be detected by the PMTs, creating a signal labeled S1. The ions are prevented from recombining by the applied electric field, which causes the electrons to drift upwards. When the electrons reach the larger field at the liquid-gas interface they are extracted into the gas phase and rapidly accelerate, producing a secondary scintillation signal labeled S2. The size of the S1 and S2 signals are proportional to the initial number of photons and ions created by the recoil event. Since electronic recoils produce larger ionization yields than nuclear recoils, the S1/S2 ratio contains information that can be used to discriminate electronic and nuclear recoils. Migdal events, however, are a combination of nuclear and electronic recoils, and thus will not necessarily resemble either. Instead, their classification will depend on the fraction of energy coming from each component.

\begin{figure}[htb]
    \centering
    \includegraphics[width=\columnwidth]{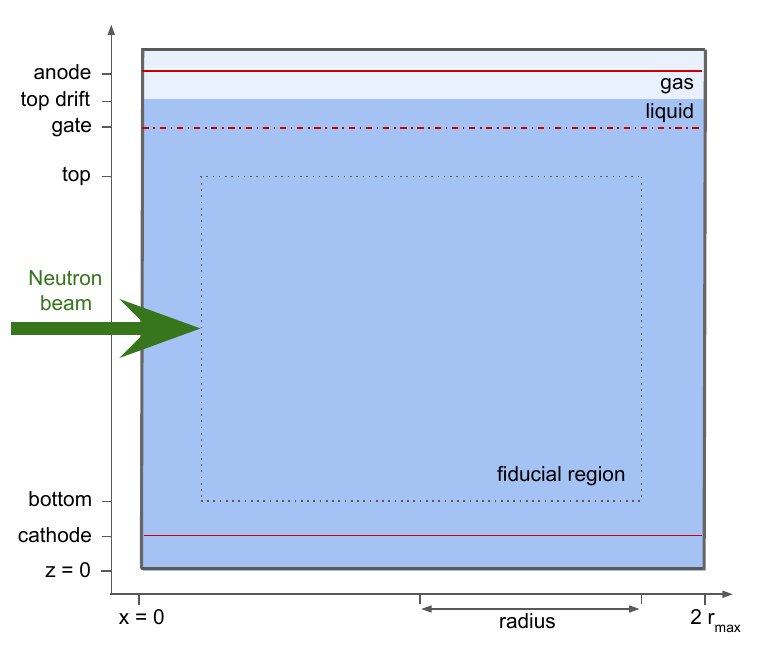}
    \caption{Cross sectional geometry of a liquid xenon TPC (not to scale), where cylindrical symmetry is assumed. Dimensions for the detector modelled in this work are given in table~\ref{tab:dimensions}.}
    \label{fig:geometry}
\end{figure}

\begin{table}[hbt]
    \caption{The dimensions of the xenon detector modelled in this work. The corresponding geometric parameters are shown in Fig.~\ref{fig:geometry}.}
    \centering
    \begin{tabular}{cc}
    \hline
    dimension & position (mm)\\
    \hline
    r$_{\mathrm{max}}$       &  120.\\
    radius      &  100.\\ 
    cathode     &  20.0\\
    bottom      &  40.0\\
    top         &  160.\\
    gate        &  190.\\
    top drift   &  195.\\
    anode       &  200.\\
    \hline
    \end{tabular}
    \label{tab:dimensions}
\end{table}

Working with the results presented in~\cite{Barbeau:2007qh}, we model our 17~keV neutron beam as having a Gaussian profile with full-width at half max of 5.9~cm and having a peak flux of 1455~neutrons/cm$^2$/s (representing a modest 11\% increase in total flux). With this in mind we model a relatively small xenon detector with dimensions given in table~\ref{tab:dimensions} and shown schematically in Fig.~\ref{fig:geometry}. The dimensions are motivated by the requirements of the neutron beam source, though we do not consider how one would couple the neutron beam to the detector. The size of the fiducial region captures $>98\%$ of the incoming neutron flux and is deep enough for $\sim63\%$ of incoming neutrons to scatter.  These dimensions result in a 10~kg fiducial region, based on a liquid xenon density of $\rho = 2.8611$~g/cm$^3$ (at 173~K). Taking the mean free path and detector geometry into account, the neutron beam provides an average flux of 100~neutrons/cm$^2$/s within the detector fiducial volume.

The detector properties, summarized in table~\ref{tab:detector}, were chosen to represent what is achievable in a xenon TPC with current technology~\cite{LUX:2012kmp}. While longer electron lifetimes are achievable (at increased expense), the chosen value is already more than 3 times the maximum drift time given the size of the detector. The drift field was chosen to maximise the electronic vs. nuclear recoil discrimination power~\cite{LUX:2020car}. Owing to their lower rates, the neutrino sources would require a scaled up detector. Larger detectors would likely have a smaller $g_1$ and require longer electron lifetimes, but otherwise the results of this detector simulation should be generally applicable to a larger detector.
 
\begin{table}[tb]
    \caption{The xenon detector properties.}
    \centering
    \begin{tabular}{cc}
    \hline
    parameter & value\\
    \hline
    $g_1$       &  0.15 PE/$\gamma$\\
    $g_2$       &  24 PE/$e^\text{-}$\\ 
    field       &  300 V/cm\\
    $e^\text{-}$ lifetime &  350.$\mu$s\\
    min S1      &  2 phd  \\
    min S2      &  250 phd\\
    no. PMTs    &  60 \\
    \hline
    \end{tabular}
    \label{tab:detector}
\end{table}

\subsection{Simulating the Migdal effect}
As outlined in section~\ref{sec:mig}, the energy deposition of a Migdal recoil has three components: the initial nuclear recoil, the ejected electron and the subsequent deexcitation of the atomic system. In previous dark matter sensitivity studies, the Migdal event is treated as a single injection of electronic energy, including the quenched nuclear recoil energy. This ignores two points: the quanta produced by the nuclear recoil are subject to fluctuations (therefore, so is the quenching factor) and the electronic process has two components. Including these points is necessary to properly model the detector response to the Migdal effect. Fluctuations in the quenching factor also have a dramatic effect on the detector response to near-threshold nuclear recoil events, as we have by design in dark matter Migdal analyses. Here, we model the entire process on an event-by-event basis using the NEST simulation code~\cite{Szydagis_2011,szydagis_m_2021_5676553}. 

The Migdal effect is incorporated into the NEST nuclear recoil event workflow through the addition of the following steps:
\begin{enumerate}
    \item After an energy has been selected from a given NR distribution, calculate the maximum allowed $E_{\rm EM}$. For an incident, non-relativistic neutron with kinetic energy $E_n = m_nv_n^2/2$ impinging upon a nuclear target of mass $m_{\rm T}$, this is 
    \bea
    E_{\rm EM, max} = \frac{\mu_{\rm T}v_{n}^2}{2} = \frac{\mu_{\rm T}E_n}{m_n},
    \eea
    where $E_n$ is related to the nuclear recoil energy through the formula in Eq.(\ref{eq:ENRneutron}).
    For an incident neutrino that scatters at an angle of $\theta_{\nu\nu'}$, one finds the total electronic energy, $\Delta E$, arising from the sum of the ionized electron plus the energy from de-excitation is
    \bea
    \Delta E \simeq \frac{E_{\nu}^2(1-\cos\theta_{\nu\nu'})-m_{\rm T}E_{\rm R}}{E_{\nu}(1-\cos\theta_{\nu\nu'}) - E_{\rm R}},    
    \eea
    where we have made the approximation that the nuclear recoil energy and the electronic energy are small compared to the target mass and neutrino energy.
    \item Loop over atomic shell and attempt to randomly ionize an electron according to the probability distribution of $E_e$ in Eq.~(\ref{eq:znl}).
    \item Calculate the charge and light yield produced from the three separate sources of energy $E_R$, $E_e$ and $E_{nl}$ and sum them.
    \item Calculate the quanta from the summed yields and then the corresponding S1 and S2 signals.
\end{enumerate}

\begin{table*}[htb]
    \caption{Comparison of the calculated and simulated ratio of Migdal events to NR events. The simulated rate and ratio is after cuts on S1 and S2.}
    \centering
    \begin{tabular}{|c|c|c|c|}
    \hline
    Source              & Calc. ratio & Sim. ratio & Sim. rate/kg/day \\
    \hline
    neutron (17~keV)    & $6.0\times 10^{-4}$ & 0.1 & $600$\\
    reactor neutrinos   & $1.7\times 10^{-4}$ & 0.1 & $4.3\times 10^{-4}$ \\ 
    SNS neutrinos       & $1.5\times 10^{-2}$ & 0.02& $8.8\times 10^{-3}$\\
    $^{51}$Cr neutrinos & $5.4\times 10^{-6}$ & $\infty$ & $8.2\times 10^{-6}$  \\
    \hline
    \end{tabular}
    \label{tab:results}
\end{table*}

Steps 3. and 4. are performed by NEST using the yield calculations for a nuclear recoil with energy $E_R$ and electronic recoils with energy $E_{nl}$ and $E_{e}$.  The electronic recoils use NEST's $\beta$ model, which we found to be most suitable for modelling a Migdal event (see appendix~\ref{app:yieldModel} for further details). Here we have assumed that we can treat the yield calculations for the three sources of energy independently, while treating the quanta jointly. While these choices may not capture the microphysics of a Migdal event, we consider this a starting point for future experimental or theoretical explorations. Indeed, the reason an experimental calibration of the Migdal effect is desired is because the detector response is unknown. This procedure can also be applied to the simulation of the Migdal effect for WIMPs\footnote{Our implementation is available at \href{https://doi.org/10.5281/zenodo.5587760}{10.5281/zenodo.5587760}}.

\begin{figure*}[tb!]
\includegraphics[angle=0,width=.4\textwidth]{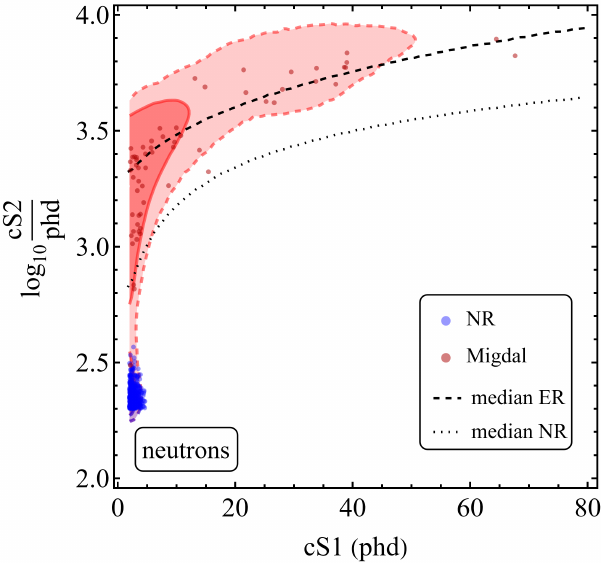}~~~~
\includegraphics[angle=0,width=.4\textwidth]{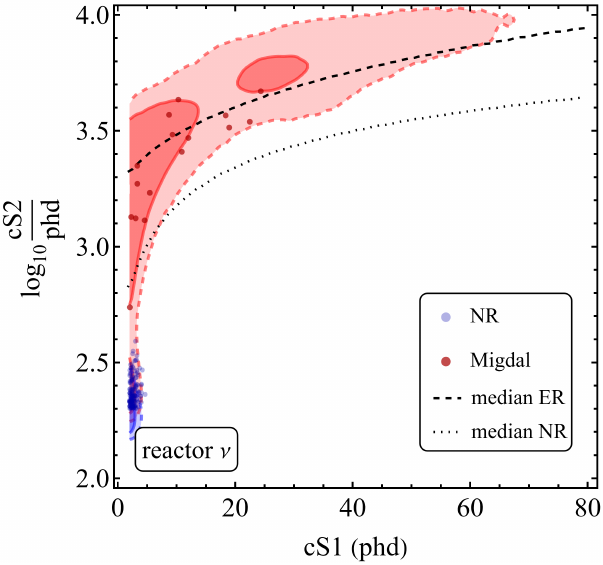}\\
\includegraphics[angle=0,width=.4\textwidth]{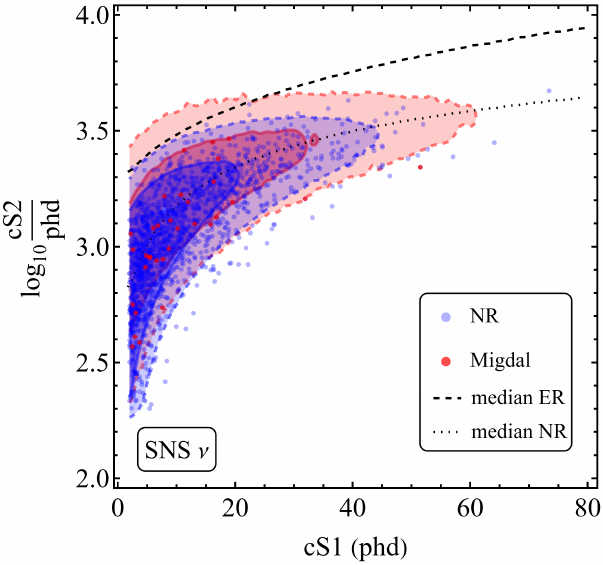}~~~~
\includegraphics[angle=0,width=.4\textwidth]{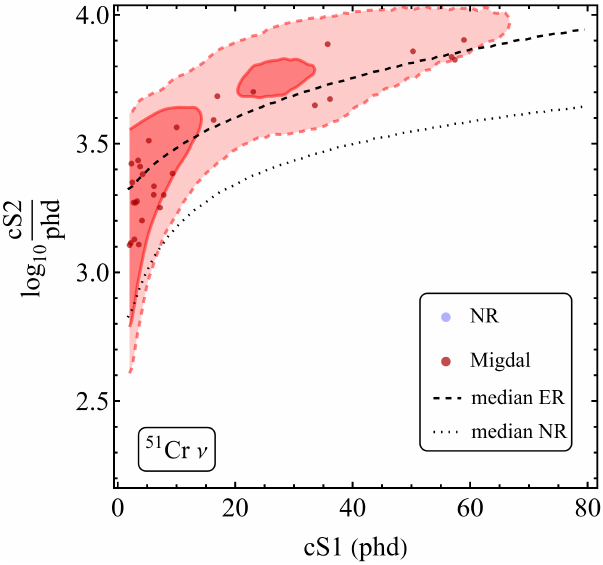}\\
\caption{The 1 and 2 $\sigma$ confidence regions for NR and Migdal events in the S1-S2 plane for the four sources considered in this work. The overlayed dots show events from a representative exposure: 1~kg-day for neutrons, and 10$^{-1}$ tonne-years for reactor neutrinos, 10$^{-2}$ tonne-years for SNS neutrinos and 10 tonne-years for chromium neutrinos.}
\label{fig:xe_S1S2}
\end{figure*}

We performed simulations following the above steps for each of our four sources of nuclear recoils: neutrons, reactor neutrinos, SNS neutrinos and the chromium decay neutrinos. In each case, we simulated a sufficient number  of events ($\sim 10^5$ - $10^6$ events) to estimate the ratio of NR to Migdal events and the absolute Migdal rate passing basic detector cuts. The results are summarized in table~\ref{tab:results}. Given the low probability of Migdal events, even in the best cases our results were dominated by nuclear recoil samples. To improve our sampling of the S1-S2 distribution of Migdal events, further simulations were performed with a scaled Migdal probability. The resulting NR and Migdal events were binned in the S1-log(S2) to produce the 1 and 2$\sigma$ confidence regions given in Fig.~\ref{fig:xe_S1S2}. These regions indicate how well separated the NR and Migdal events are, but do not demonstrate how well they can be distinguished on an event-by-event basis. To address this we overlay a representative sample of events at the true NR to Migdal event ratio.

The distribution of Migdal events in S1-log(S2) is well separated from the NR distribution for three of the sources: the neutrons, reactor neutrinos and chromium source. This is due to the low energy of the NR from each of these sources, which only provides an insignificant contribution to the observed signal for most Migdal events. For each of these sources the Migdal distribution sits slightly above the median electronic recoil curve due to the fact that we have added the yields of two electronic recoil recoil events (one for the ejected Migdal electron and one for the de-excitation process). Whether this is a physical feature remains to be seen. The neutron Migdal distribution exhibits a second island in the 1$\sigma$ confidence region at low S1 and S2 (hidden by the NR in Fig.~\ref{fig:xe_S1S2}), while the chromium distribution exhibits one at higher S1 and S2. The reactor Migdal distribution exhibits both of these extra islands, for a total of three. These islands correspond to the different atomic shells from which the Migdal electrons were ejected. For these three sources, the most commonly observed (above threshold) Migdal events come from the $n = 4$ shell, which produces the main cluster of events around S1 $<$ 10 phd and $\log{\frac{S2}{\text{phd}}}=  3$ - 3.5. The $n = 3$ ($n = 5$) shell has more (less) binding energy and so it appears at higher (lower) S1 and S2. 

The shape of the NR spectrum has an effect on the relative rates of Migdal events from each the atomic shells, causing the observed differences in the chromium, reactor and neutron sources. The similar NR spectra from neutrons and reactor neutrinos causes these two sources to have a similar distribution and the same ratio of Migdal to NR events passing our cuts (at very different absolute rates however). The energy from nuclear recoils of the reactor and neutron sources contributes to the Migdal events from the $n = 5$ shell, pushing some of them above the S1 and S2 thresholds. This is contrasted with the chromium source, which sees very few events from the $n=5$ shell. For the SNS neutrino source, the larger energy of the NR contributes and even dominates the Migdal event's energy. This erases any structure in the Migdal event distribution and pulls the distribution down to the median NR curve.


\section{Conclusions}

In this paper, we investigated the feasibility of using liquid xenon and argon TPCs to detect Migdal events arising due to nuclear recoils from four different sources: neutrons, reactor neutrinos, pion DAR neutrinos (from the SNS) and chromium-51 decay neutrinos. We found that the Migdal rates (per unit target mass) are similar in the argon and xenon detectors even though the nuclear recoil rates are smaller in argon. This characteristic, along with the smaller neutron capture cross section, may make argon a more desirable target for Migdal studies. However, this is highly dependent on what electronic recoil threshold can be achieved in argon.

We also find that, given the small ratio of Migdal to nuclear recoil events, it is imperative that the detector has excellent nuclear/electronic recoil discrimination. This is best achieved if the nuclear recoil energies are small and so is highly dependent on the source used. Our xenon detector simulations show that the lack of separation of Migdal and NR events in S1-S2 for the SNS neutrinos makes identification impossible on an event by event basis. However, the Migdal effect does provide a $2\%$ level correction to the observed event rate from SNS neutrinos. While there is good separation of Migdal and NR events for the chromium and reactor neutrino sources, the absolute rate that passes our cuts is very low. The neutron source exhibits good separation between the Migdal and NR events at $E_n = 17$~keV. The ability to discriminate the events would be diminished as the neutron energy, and thus nuclear recoil energy, is increased.

Our results indicate that observing the Migdal effect due to neutrino scattering would be incredibly challenging. Exposures of 10-1,000~tonne-days would be required just to obtain a handful of events, before considerations of external backgrounds are taken into account. These exposures are not realistic for the 10~kg xenon detector explored in this work. In terms of event rate, the only feasible option is the neutron source which can induce over thousands of Migdal events per day. Low-energy neutron sources thus provide an opportunity to calibrate the Migdal effect directly inside a dark matter style liquid noble TPC, in a kinematic regime that is directly analogous to low-mass dark matter scattering.

\section*{Acknowledgements} JLN and NFB were supported by the Australian Research Council through the ARC Centre of Excellence for Dark Matter Particle Physics, CE200100008. JBD acknowledges support from the National Science Foundation under grants no. PHY-1820801 and PHY-1748958, and RFL from grant PHY-2112803. ACR was supported by an Australian Government Research Training Program Scholarship.

\appendix
\section{Differential rates for the Migdal effect}
\label{app:rates}
\begin{figure*}[hbt]
\includegraphics[width=.4\textwidth]{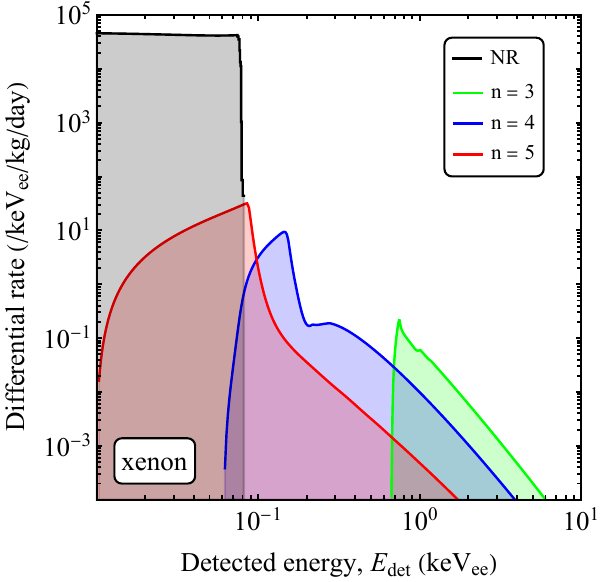}~~~
\includegraphics[width=.4\textwidth]{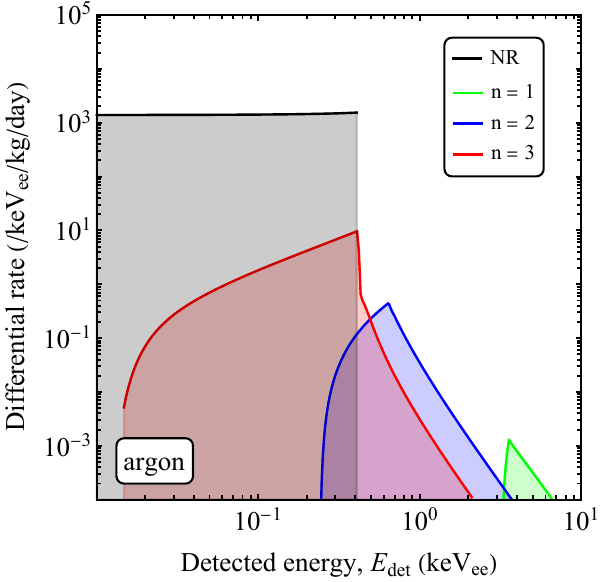}
\caption{The differential nuclear recoil and Migdal scattering rates for 17~keV neutrons as a function of detected energy. The Migdal rate is given separately for each atomic shell, $n$.}
\label{fig:neutron_rate}
\end{figure*}

\begin{figure*}[hbt]
\includegraphics[width=.37\textwidth]{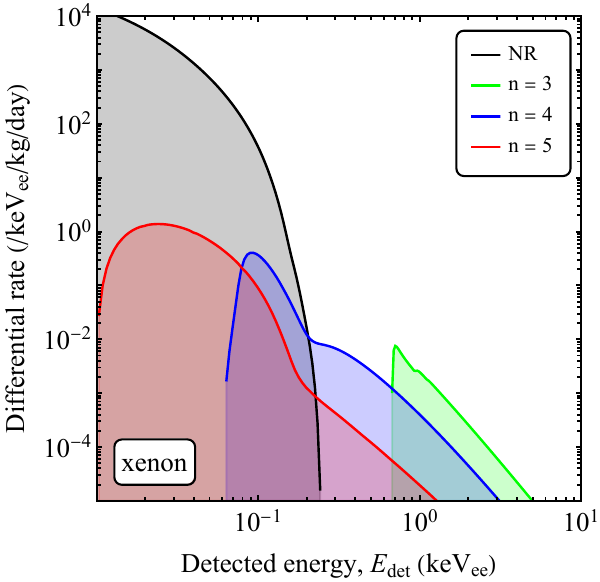}~~~
\includegraphics[width=.37\textwidth]{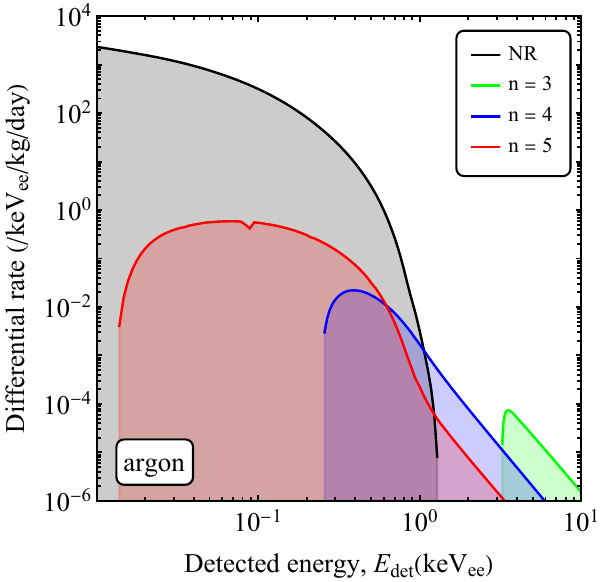}
\caption{The differential nuclear recoil and Migdal scattering rates for reactor neutrinos as a function of detected energy. The Migdal rate is given separately for each atomic shell, $n$.}
\label{fig:reactorNuMig}
\end{figure*}

\begin{figure*}[hbt]
\includegraphics[width=.37\textwidth]{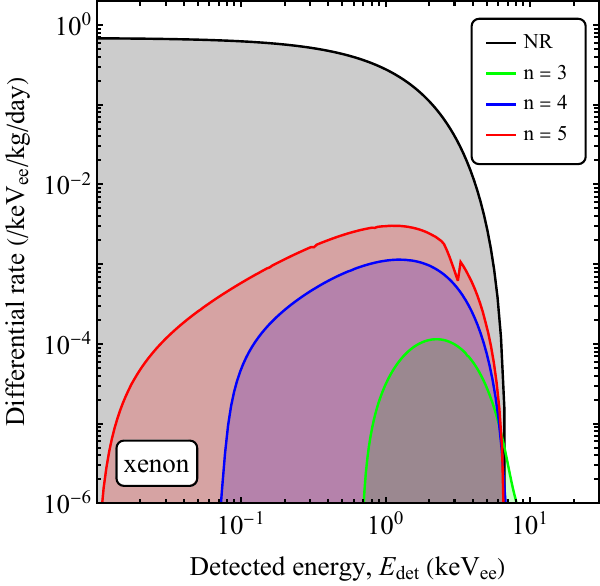}~~~ 
\includegraphics[width=.37\textwidth]{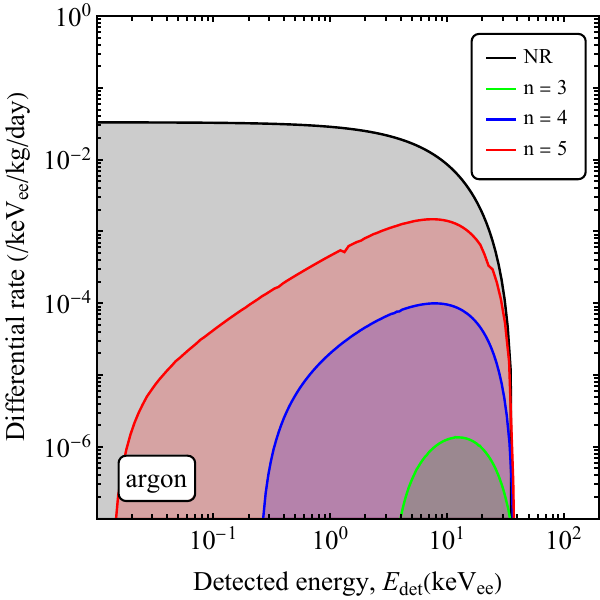} 
\caption{The differential nuclear recoil and Migdal scattering rates for SNS neutrinos as a function of detected energy. The Migdal rate is given separately for each atomic shell, $n$.}
\label{fig:snsNuMig}
\end{figure*}

\begin{figure*}[hbt]
\includegraphics[width=.37\textwidth]{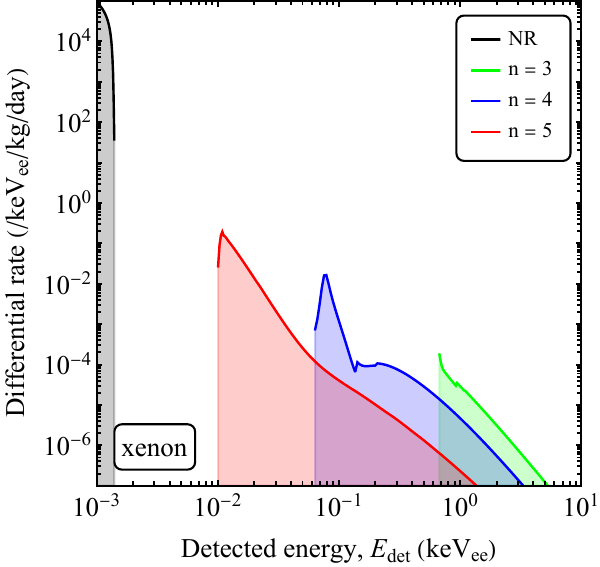}~~~ 
\includegraphics[width=.37\textwidth]{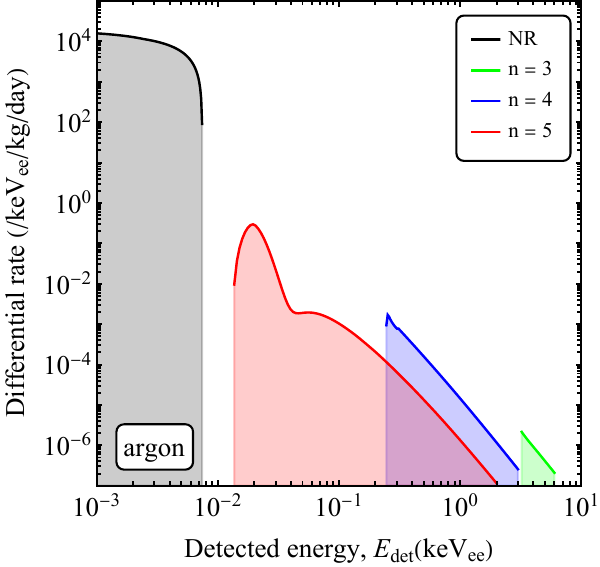} 
\caption{The differential nuclear recoil and Migdal scattering rates for chromium neutrinos as a function of detected energy. The Migdal rate is given separately for each atomic shell, $n$.}
\label{fig:crNuMig}
\end{figure*}

Figures~\ref{fig:neutron_rate}-\ref{fig:crNuMig} show the differential rates in xenon and argon for nuclear recoils and the Migdal effect due to scattering of neutrons and the three neutrino sources. These rates assume a constant quenching factor of $\mathcal{L}_{\rm Xe} = 0.15$ and $\mathcal{L}_{\rm Ar} = 0.25$. These values are derived from the low energy region ($E_R \sim .5$ keV) of lindhard theory. It has been shown that using a variable quenching factor from Lindhard theory has a minimal effect on the resulting Migdal rate~\cite{Bell:2021zkr}.

\clearpage
\section{Intrinsic backgrounds from neutron capture}
\label{app:backgrounds}
\begin{figure}[htb!]
\includegraphics[angle=0,width=.4\textwidth]{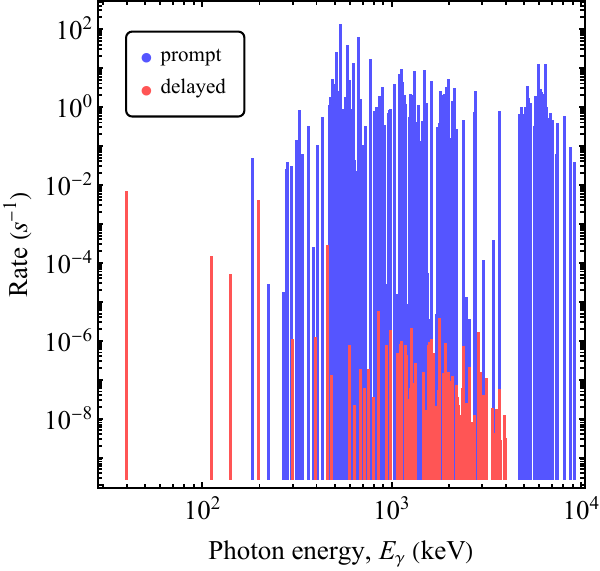}\\
\includegraphics[angle=0,width=.4\textwidth]{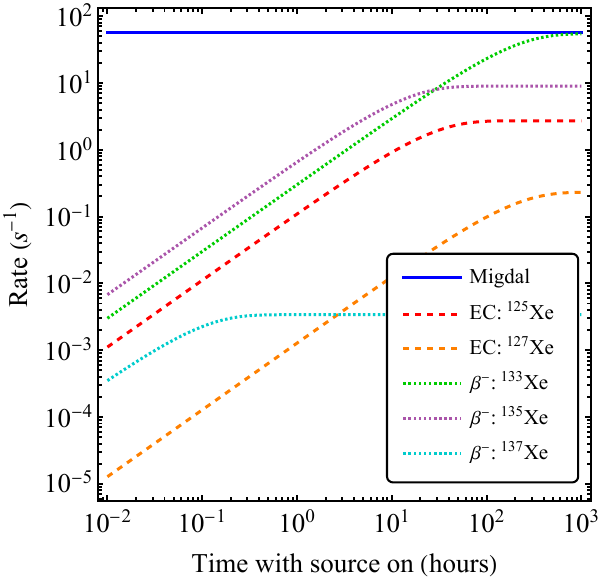}
\caption{Top: the rate of prompt (blue) and delayed (red) gamma-ray emission following neutron capture in xenon, data from~\cite{agency_database_2007}. Bottom: the rate of electron capture and $\beta$ decays of unstable xenon isotopes that accumulate over time due to neutron capture.}
\label{fig:xeGammaSpectra}
\end{figure}

\begin{figure}[htb]
\includegraphics[angle=0,width=.4\textwidth]{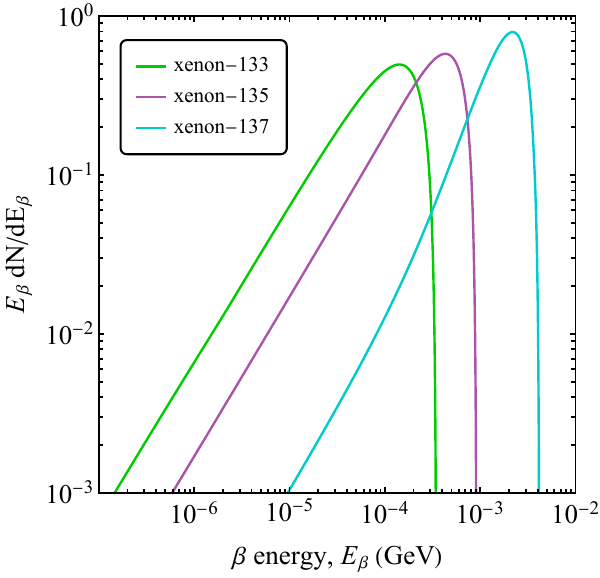}\\
\includegraphics[angle=0,width=.4\textwidth]{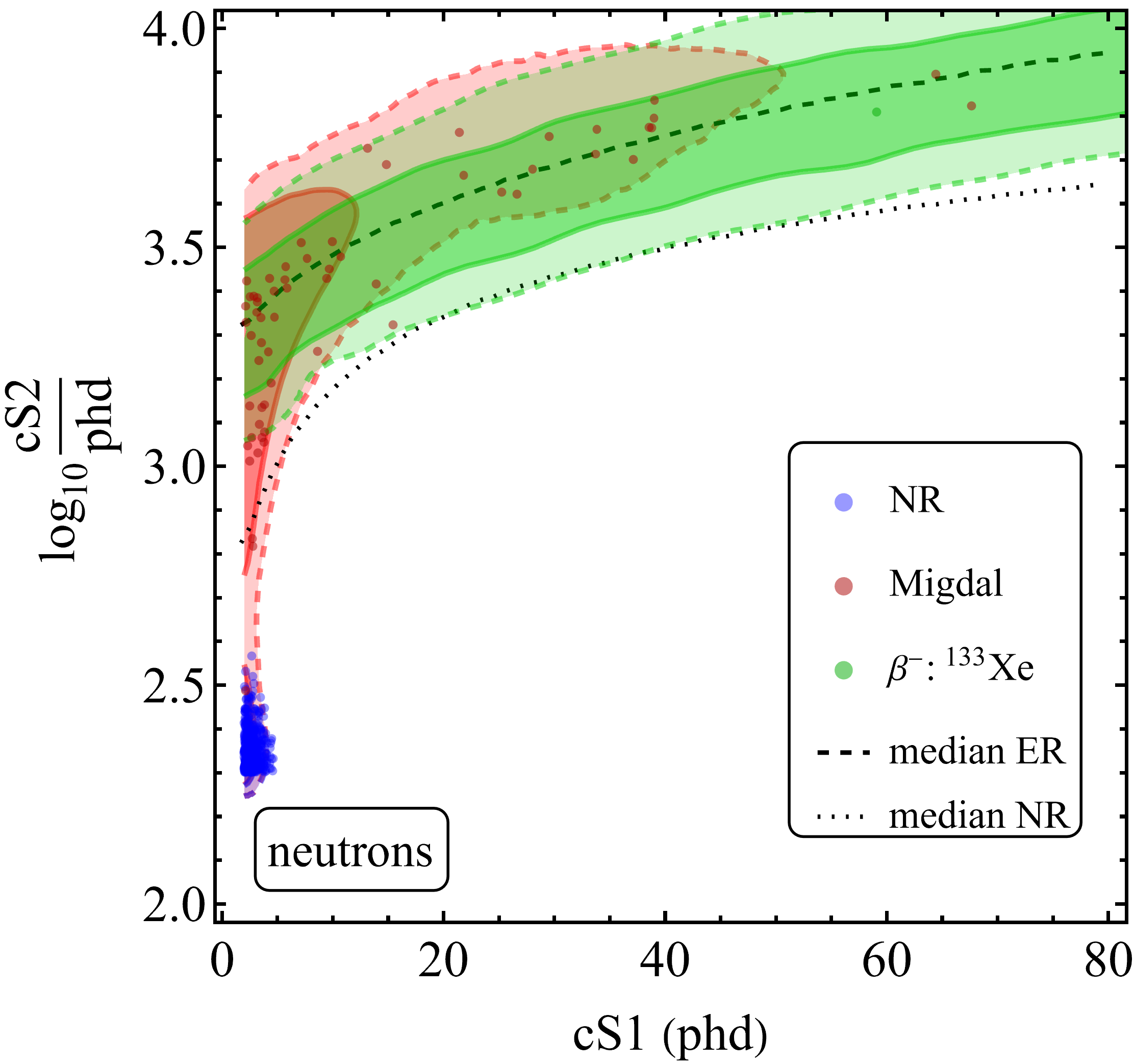}
\caption{Top: the $\beta$ decay spectra from the three relevant xenon isotopes, data from~\cite{PhysRevC.91.055504}. Bottom: the simulated S1-log(S2) distribution of $^{133}$Xe $\beta$ decay events based on 1~kg-day of exposure at the steady state background rate (after $\sim$~20~days of continuous beam).}
\label{fig:xeBetaSpectra}
\end{figure}

Radiative neutron capture rates were calculated for a neutron flux of 100~cm$^2$/s and neutron energy of 17~keV, using the capture cross section for each isotope from~\cite{JENDL4.0:2011}. The partial cross sections for each gamma transition were obtained from~\cite{agency_database_2007}, which are evaluated at $E_n = 0.025$ eV. These cross sections were used to calculate the branching fraction for each gamma transition that was assumed to be the same at $E_n = 17$~keV. The resulting spectrum of gamma rays is given in Fig.~\ref{fig:xeGammaSpectra}.
The unstable isotopes created through neutron capture then decay via $\beta$ decay and electron capture. Given the half-lives of these isotopes we assume that the xenon in the detector has time to circulate and homogeneously distribute the unstable isotopes. Our decay rates therefore include a factor of $1/2$ to account for the fact that only a fraction of the xenon will be in the fiducial/active region of the detector. The $\beta$ decay spectrum was simulated using the detector configuration in Sec.~\ref{sec:detSim} and the resulting S1-log(S2) distribution is shown in Fig.~\ref{fig:xeBetaSpectra} (right). The fraction of the $\beta$ events that have S1$<$ 50 is: $5\%$, $1\%$ and $0.1\%$ for $^{133}$Xe, $^{135}$Xe and $^{137}$Xe respectively. The dominant contribution comes from $^{133}$Xe, which has an absolute rate equivalent to the Migdal rate once steady state is reached. However, even then only a few background events are expected in the Migdal region per~kg-day.

\section{Yields for Migdal events}
\label{app:yieldModel}
\begin{figure}[htb]
\includegraphics[angle=0,width=.4\textwidth]{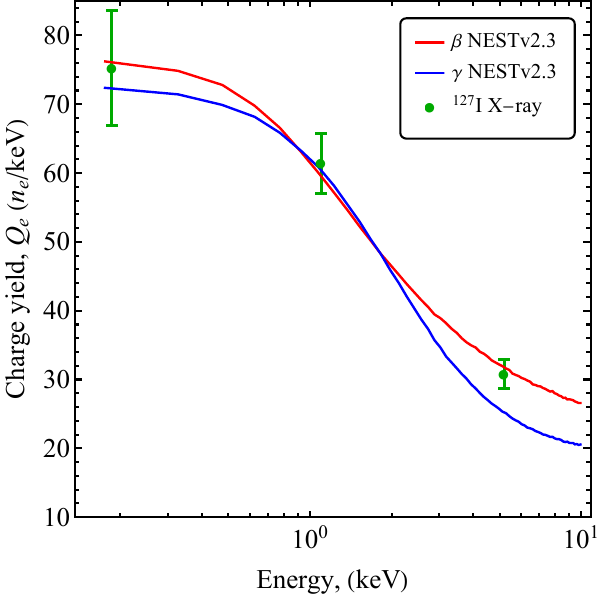}
\caption{Comparison of $\gamma$ (red) and $\beta$ (blue) charge yield at $V=180$ V/cm in NESTv2.3 with data from LUX's electron capture calibration (green)~\cite{LUX:2017ojt}. Errors correspond to statistical and systematic errors summed in quadrature.}
\label{fig:Yields}
\end{figure}

The deexcitation of a xenon atom following the ejection of a non-valence electron (in this case due to the Migdal effect) can proceed via emission of an x-ray, Auger decay or some combination. Here we use electron capture decays of $^{127}$Xe and subsequent deexcitation of $^{127}$I to inform our choice of yield model. Fig.~\ref{fig:Yields} shows that the $\beta$ model in NESTv2.3 does well at reproducing the data from LUX's electron-capture calibration~\cite{LUX:2017ojt}. While there aren't many data points to compare with the model, we are only interested in how the models perform in the neighborhood of points shown, since they correspond to electron capture from the $n=4,3,2$ shells (from left to right).

\clearpage
\bibliography{neutron_mig}

\begin{thebibliography}{43}%
\makeatletter
\providecommand \@ifxundefined [1]{%
 \@ifx{#1\undefined}
}%
\providecommand \@ifnum [1]{%
 \ifnum #1\expandafter \@firstoftwo
 \else \expandafter \@secondoftwo
 \fi
}%
\providecommand \@ifx [1]{%
 \ifx #1\expandafter \@firstoftwo
 \else \expandafter \@secondoftwo
 \fi
}%
\providecommand \natexlab [1]{#1}%
\providecommand \enquote  [1]{``#1''}%
\providecommand \bibnamefont  [1]{#1}%
\providecommand \bibfnamefont [1]{#1}%
\providecommand \citenamefont [1]{#1}%
\providecommand \href@noop [0]{\@secondoftwo}%
\providecommand \href [0]{\begingroup \@sanitize@url \@href}%
\providecommand \@href[1]{\@@startlink{#1}\@@href}%
\providecommand \@@href[1]{\endgroup#1\@@endlink}%
\providecommand \@sanitize@url [0]{\catcode `\\12\catcode `\$12\catcode
  `\&12\catcode `\#12\catcode `\^12\catcode `\_12\catcode `\%12\relax}%
\providecommand \@@startlink[1]{}%
\providecommand \@@endlink[0]{}%
\providecommand \url  [0]{\begingroup\@sanitize@url \@url }%
\providecommand \@url [1]{\endgroup\@href {#1}{\urlprefix }}%
\providecommand \urlprefix  [0]{URL }%
\providecommand \Eprint [0]{\href }%
\providecommand \doibase [0]{http://dx.doi.org/}%
\providecommand \selectlanguage [0]{\@gobble}%
\providecommand \bibinfo  [0]{\@secondoftwo}%
\providecommand \bibfield  [0]{\@secondoftwo}%
\providecommand \translation [1]{[#1]}%
\providecommand \BibitemOpen [0]{}%
\providecommand \bibitemStop [0]{}%
\providecommand \bibitemNoStop [0]{.\EOS\space}%
\providecommand \EOS [0]{\spacefactor3000\relax}%
\providecommand \BibitemShut  [1]{\csname bibitem#1\endcsname}%
\let\auto@bib@innerbib\@empty
\bibitem [{\citenamefont {Migdal}(1941)}]{Migdal:1941}%
  \BibitemOpen
  \bibfield  {author} {\bibinfo {author} {\bibfnamefont {A.}~\bibnamefont
  {Migdal}},\ }\href@noop {} {\bibfield  {journal} {\bibinfo  {journal}
  {J.Phys.(USSR)}\ }\textbf {\bibinfo {volume} {4}},\ \bibinfo {pages} {449}
  (\bibinfo {year} {1941})}\BibitemShut {NoStop}%
\bibitem [{\citenamefont {Rapaport}\ \emph {et~al.}(1975)\citenamefont
  {Rapaport}, \citenamefont {Asaro},\ and\ \citenamefont
  {Perlman}}]{PhysRevC.11.1740}%
  \BibitemOpen
  \bibfield  {author} {\bibinfo {author} {\bibfnamefont {M.~S.}\ \bibnamefont
  {Rapaport}}, \bibinfo {author} {\bibfnamefont {F.}~\bibnamefont {Asaro}}, \
  and\ \bibinfo {author} {\bibfnamefont {I.}~\bibnamefont {Perlman}},\ }\href
  {\doibase 10.1103/PhysRevC.11.1740} {\bibfield  {journal} {\bibinfo
  {journal} {Phys. Rev. C}\ }\textbf {\bibinfo {volume} {11}},\ \bibinfo
  {pages} {1740} (\bibinfo {year} {1975})}\BibitemShut {NoStop}%
\bibitem [{\citenamefont {Couratin}\ \emph {et~al.}(2012)\citenamefont
  {Couratin}, \citenamefont {Velten}, \citenamefont {Fl\'echard}, \citenamefont
  {Li\'enard}, \citenamefont {Ban}, \citenamefont {Cassimi}, \citenamefont
  {Delahaye}, \citenamefont {Durand}, \citenamefont {Hennecart}, \citenamefont
  {Mauger}, \citenamefont {M\'ery}, \citenamefont {Naviliat-Cuncic},
  \citenamefont {Patyk}, \citenamefont {Rodr\'{\i}guez}, \citenamefont
  {Siegie\ifmmode \acute{n}\else~\'{n}\fi{} Iwaniuk},\ and\ \citenamefont
  {Thomas}}]{PhysRevLett.108.243201}%
  \BibitemOpen
  \bibfield  {author} {\bibinfo {author} {\bibfnamefont {C.}~\bibnamefont
  {Couratin}}, \bibinfo {author} {\bibfnamefont {P.}~\bibnamefont {Velten}},
  \bibinfo {author} {\bibfnamefont {X.}~\bibnamefont {Fl\'echard}}, \bibinfo
  {author} {\bibfnamefont {E.}~\bibnamefont {Li\'enard}}, \bibinfo {author}
  {\bibfnamefont {G.}~\bibnamefont {Ban}}, \bibinfo {author} {\bibfnamefont
  {A.}~\bibnamefont {Cassimi}}, \bibinfo {author} {\bibfnamefont
  {P.}~\bibnamefont {Delahaye}}, \bibinfo {author} {\bibfnamefont
  {D.}~\bibnamefont {Durand}}, \bibinfo {author} {\bibfnamefont
  {D.}~\bibnamefont {Hennecart}}, \bibinfo {author} {\bibfnamefont
  {F.}~\bibnamefont {Mauger}}, \bibinfo {author} {\bibfnamefont
  {A.}~\bibnamefont {M\'ery}}, \bibinfo {author} {\bibfnamefont
  {O.}~\bibnamefont {Naviliat-Cuncic}}, \bibinfo {author} {\bibfnamefont
  {Z.}~\bibnamefont {Patyk}}, \bibinfo {author} {\bibfnamefont
  {D.}~\bibnamefont {Rodr\'{\i}guez}}, \bibinfo {author} {\bibfnamefont
  {K.}~\bibnamefont {Siegie\ifmmode \acute{n}\else~\'{n}\fi{} Iwaniuk}}, \ and\
  \bibinfo {author} {\bibfnamefont {J.-C.}\ \bibnamefont {Thomas}},\ }\href
  {\doibase 10.1103/PhysRevLett.108.243201} {\bibfield  {journal} {\bibinfo
  {journal} {Phys. Rev. Lett.}\ }\textbf {\bibinfo {volume} {108}},\ \bibinfo
  {pages} {243201} (\bibinfo {year} {2012})}\BibitemShut {NoStop}%
\bibitem [{\citenamefont {Vergados}\ and\ \citenamefont
  {Ejiri}(2005)}]{Vergados:2004bm}%
  \BibitemOpen
  \bibfield  {author} {\bibinfo {author} {\bibfnamefont {J.~D.}\ \bibnamefont
  {Vergados}}\ and\ \bibinfo {author} {\bibfnamefont {H.}~\bibnamefont
  {Ejiri}},\ }\href {\doibase 10.1016/j.physletb.2004.11.085} {\bibfield
  {journal} {\bibinfo  {journal} {Phys. Lett.}\ }\textbf {\bibinfo {volume}
  {B606}},\ \bibinfo {pages} {313} (\bibinfo {year} {2005})},\ \Eprint
  {http://arxiv.org/abs/hep-ph/0401151} {arXiv:hep-ph/0401151 [hep-ph]}
  \BibitemShut {NoStop}%
\bibitem [{\citenamefont {Ejiri}\ \emph {et~al.}(2006)\citenamefont {Ejiri},
  \citenamefont {Moustakidis},\ and\ \citenamefont {Vergados}}]{Ejiri:2005aj}%
  \BibitemOpen
  \bibfield  {author} {\bibinfo {author} {\bibfnamefont {H.}~\bibnamefont
  {Ejiri}}, \bibinfo {author} {\bibfnamefont {C.~C.}\ \bibnamefont
  {Moustakidis}}, \ and\ \bibinfo {author} {\bibfnamefont {J.~D.}\ \bibnamefont
  {Vergados}},\ }\href {\doibase 10.1016/j.physletb.2006.03.037} {\bibfield
  {journal} {\bibinfo  {journal} {Phys. Lett.}\ }\textbf {\bibinfo {volume}
  {B639}},\ \bibinfo {pages} {218} (\bibinfo {year} {2006})},\ \Eprint
  {http://arxiv.org/abs/hep-ph/0510042} {arXiv:hep-ph/0510042 [hep-ph]}
  \BibitemShut {NoStop}%
\bibitem [{\citenamefont {Moustakidis}\ \emph {et~al.}(2005)\citenamefont
  {Moustakidis}, \citenamefont {Vergados},\ and\ \citenamefont
  {Ejiri}}]{Moustakidis:2005gx}%
  \BibitemOpen
  \bibfield  {author} {\bibinfo {author} {\bibfnamefont {C.~C.}\ \bibnamefont
  {Moustakidis}}, \bibinfo {author} {\bibfnamefont {J.~D.}\ \bibnamefont
  {Vergados}}, \ and\ \bibinfo {author} {\bibfnamefont {H.}~\bibnamefont
  {Ejiri}},\ }\href {\doibase 10.1016/j.nuclphysb.2005.08.033} {\bibfield
  {journal} {\bibinfo  {journal} {Nucl. Phys.}\ }\textbf {\bibinfo {volume}
  {B727}},\ \bibinfo {pages} {406} (\bibinfo {year} {2005})},\ \Eprint
  {http://arxiv.org/abs/hep-ph/0507123} {arXiv:hep-ph/0507123 [hep-ph]}
  \BibitemShut {NoStop}%
\bibitem [{\citenamefont {Bernabei}\ \emph {et~al.}(2008)\citenamefont
  {Bernabei} \emph {et~al.}}]{Bernabei:2007gr}%
  \BibitemOpen
  \bibfield  {author} {\bibinfo {author} {\bibfnamefont {R.}~\bibnamefont
  {Bernabei}} \emph {et~al.},\ }\href {\doibase 10.1103/PhysRevD.77.023506}
  {\bibfield  {journal} {\bibinfo  {journal} {Phys. Rev.}\ }\textbf {\bibinfo
  {volume} {D77}},\ \bibinfo {pages} {023506} (\bibinfo {year} {2008})},\
  \Eprint {http://arxiv.org/abs/0712.0562} {arXiv:0712.0562 [astro-ph]}
  \BibitemShut {NoStop}%
\bibitem [{\citenamefont {Vergados}\ \emph {et~al.}(2013)\citenamefont
  {Vergados}, \citenamefont {Ejiri},\ and\ \citenamefont
  {Savvidy}}]{Vergados:2013raa}%
  \BibitemOpen
  \bibfield  {author} {\bibinfo {author} {\bibfnamefont {J.~D.}\ \bibnamefont
  {Vergados}}, \bibinfo {author} {\bibfnamefont {H.}~\bibnamefont {Ejiri}}, \
  and\ \bibinfo {author} {\bibfnamefont {K.~G.}\ \bibnamefont {Savvidy}},\
  }\href {\doibase 10.1016/j.nuclphysb.2013.09.010} {\bibfield  {journal}
  {\bibinfo  {journal} {Nucl. Phys.}\ }\textbf {\bibinfo {volume} {B877}},\
  \bibinfo {pages} {36} (\bibinfo {year} {2013})},\ \Eprint
  {http://arxiv.org/abs/1307.4713} {arXiv:1307.4713 [hep-ph]} \BibitemShut
  {NoStop}%
\bibitem [{\citenamefont {Ibe}\ \emph {et~al.}(2018)\citenamefont {Ibe},
  \citenamefont {Nakano}, \citenamefont {Shoji},\ and\ \citenamefont
  {Suzuki}}]{Ibe:2017yqa}%
  \BibitemOpen
  \bibfield  {author} {\bibinfo {author} {\bibfnamefont {M.}~\bibnamefont
  {Ibe}}, \bibinfo {author} {\bibfnamefont {W.}~\bibnamefont {Nakano}},
  \bibinfo {author} {\bibfnamefont {Y.}~\bibnamefont {Shoji}}, \ and\ \bibinfo
  {author} {\bibfnamefont {K.}~\bibnamefont {Suzuki}},\ }\href {\doibase
  10.1007/JHEP03(2018)194} {\bibfield  {journal} {\bibinfo  {journal} {JHEP}\
  }\textbf {\bibinfo {volume} {03}},\ \bibinfo {pages} {194} (\bibinfo {year}
  {2018})},\ \Eprint {http://arxiv.org/abs/1707.07258} {arXiv:1707.07258
  [hep-ph]} \BibitemShut {NoStop}%
\bibitem [{\citenamefont {Dolan}\ \emph {et~al.}(2017)\citenamefont {Dolan},
  \citenamefont {Kahlhoefer},\ and\ \citenamefont {McCabe}}]{Dolan:2017xbu}%
  \BibitemOpen
  \bibfield  {author} {\bibinfo {author} {\bibfnamefont {M.~J.}\ \bibnamefont
  {Dolan}}, \bibinfo {author} {\bibfnamefont {F.}~\bibnamefont {Kahlhoefer}}, \
  and\ \bibinfo {author} {\bibfnamefont {C.}~\bibnamefont {McCabe}},\
  }\href@noop {} {\  (\bibinfo {year} {2017})},\ \Eprint
  {http://arxiv.org/abs/1711.09906} {arXiv:1711.09906 [hep-ph]} \BibitemShut
  {NoStop}%
\bibitem [{\citenamefont {Bell}\ \emph {et~al.}(2020)\citenamefont {Bell},
  \citenamefont {Dent}, \citenamefont {Newstead}, \citenamefont {Sabharwal},\
  and\ \citenamefont {Weiler}}]{Bell:2019egg}%
  \BibitemOpen
  \bibfield  {author} {\bibinfo {author} {\bibfnamefont {N.~F.}\ \bibnamefont
  {Bell}}, \bibinfo {author} {\bibfnamefont {J.~B.}\ \bibnamefont {Dent}},
  \bibinfo {author} {\bibfnamefont {J.~L.}\ \bibnamefont {Newstead}}, \bibinfo
  {author} {\bibfnamefont {S.}~\bibnamefont {Sabharwal}}, \ and\ \bibinfo
  {author} {\bibfnamefont {T.~J.}\ \bibnamefont {Weiler}},\ }\href {\doibase
  10.1103/PhysRevD.101.015012} {\bibfield  {journal} {\bibinfo  {journal}
  {Phys. Rev. D}\ }\textbf {\bibinfo {volume} {101}},\ \bibinfo {pages}
  {015012} (\bibinfo {year} {2020})},\ \Eprint
  {http://arxiv.org/abs/1905.00046} {arXiv:1905.00046 [hep-ph]} \BibitemShut
  {NoStop}%
\bibitem [{\citenamefont {Baxter}\ \emph {et~al.}(2020)\citenamefont {Baxter},
  \citenamefont {Kahn},\ and\ \citenamefont {Krnjaic}}]{Baxter:2019pnz}%
  \BibitemOpen
  \bibfield  {author} {\bibinfo {author} {\bibfnamefont {D.}~\bibnamefont
  {Baxter}}, \bibinfo {author} {\bibfnamefont {Y.}~\bibnamefont {Kahn}}, \ and\
  \bibinfo {author} {\bibfnamefont {G.}~\bibnamefont {Krnjaic}},\ }\href
  {\doibase 10.1103/PhysRevD.101.076014} {\bibfield  {journal} {\bibinfo
  {journal} {Phys. Rev. D}\ }\textbf {\bibinfo {volume} {101}},\ \bibinfo
  {pages} {076014} (\bibinfo {year} {2020})},\ \Eprint
  {http://arxiv.org/abs/1908.00012} {arXiv:1908.00012 [hep-ph]} \BibitemShut
  {NoStop}%
\bibitem [{\citenamefont {Essig}\ \emph {et~al.}(2020)\citenamefont {Essig},
  \citenamefont {Pradler}, \citenamefont {Sholapurkar},\ and\ \citenamefont
  {Yu}}]{Essig:2019xkx}%
  \BibitemOpen
  \bibfield  {author} {\bibinfo {author} {\bibfnamefont {R.}~\bibnamefont
  {Essig}}, \bibinfo {author} {\bibfnamefont {J.}~\bibnamefont {Pradler}},
  \bibinfo {author} {\bibfnamefont {M.}~\bibnamefont {Sholapurkar}}, \ and\
  \bibinfo {author} {\bibfnamefont {T.-T.}\ \bibnamefont {Yu}},\ }\href
  {\doibase 10.1103/PhysRevLett.124.021801} {\bibfield  {journal} {\bibinfo
  {journal} {Phys. Rev. Lett.}\ }\textbf {\bibinfo {volume} {124}},\ \bibinfo
  {pages} {021801} (\bibinfo {year} {2020})},\ \Eprint
  {http://arxiv.org/abs/1908.10881} {arXiv:1908.10881 [hep-ph]} \BibitemShut
  {NoStop}%
\bibitem [{\citenamefont {Akerib}\ \emph {et~al.}(2019)\citenamefont {Akerib}
  \emph {et~al.}}]{Akerib:2018hck}%
  \BibitemOpen
  \bibfield  {author} {\bibinfo {author} {\bibfnamefont {D.~S.}\ \bibnamefont
  {Akerib}} \emph {et~al.} (\bibinfo {collaboration} {LUX}),\ }\href {\doibase
  10.1103/PhysRevLett.122.131301} {\bibfield  {journal} {\bibinfo  {journal}
  {Phys. Rev. Lett.}\ }\textbf {\bibinfo {volume} {122}},\ \bibinfo {pages}
  {131301} (\bibinfo {year} {2019})},\ \Eprint
  {http://arxiv.org/abs/1811.11241} {arXiv:1811.11241 [astro-ph.CO]}
  \BibitemShut {NoStop}%
\bibitem [{\citenamefont {Armengaud}\ \emph {et~al.}(2019)\citenamefont
  {Armengaud} \emph {et~al.}}]{Armengaud:2019kfj}%
  \BibitemOpen
  \bibfield  {author} {\bibinfo {author} {\bibfnamefont {E.}~\bibnamefont
  {Armengaud}} \emph {et~al.} (\bibinfo {collaboration} {EDELWEISS}),\ }\href
  {\doibase 10.1103/PhysRevD.99.082003} {\bibfield  {journal} {\bibinfo
  {journal} {Phys. Rev. D}\ }\textbf {\bibinfo {volume} {99}},\ \bibinfo
  {pages} {082003} (\bibinfo {year} {2019})},\ \Eprint
  {http://arxiv.org/abs/1901.03588} {arXiv:1901.03588 [astro-ph.GA]}
  \BibitemShut {NoStop}%
\bibitem [{\citenamefont {Aprile}\ \emph {et~al.}(2019)\citenamefont {Aprile}
  \emph {et~al.}}]{Aprile:2019jmx}%
  \BibitemOpen
  \bibfield  {author} {\bibinfo {author} {\bibfnamefont {E.}~\bibnamefont
  {Aprile}} \emph {et~al.} (\bibinfo {collaboration} {XENON}),\ }\href
  {\doibase 10.1103/PhysRevLett.123.241803} {\bibfield  {journal} {\bibinfo
  {journal} {Phys. Rev. Lett.}\ }\textbf {\bibinfo {volume} {123}},\ \bibinfo
  {pages} {241803} (\bibinfo {year} {2019})},\ \Eprint
  {http://arxiv.org/abs/1907.12771} {arXiv:1907.12771 [hep-ex]} \BibitemShut
  {NoStop}%
\bibitem [{\citenamefont {Wang}\ \emph {et~al.}(2020)\citenamefont {Wang} \emph
  {et~al.}}]{Wang:2019wwo}%
  \BibitemOpen
  \bibfield  {author} {\bibinfo {author} {\bibfnamefont {Y.}~\bibnamefont
  {Wang}} \emph {et~al.} (\bibinfo {collaboration} {CDEX}),\ }\href {\doibase
  10.1103/PhysRevD.101.052003} {\bibfield  {journal} {\bibinfo  {journal}
  {Phys. Rev. D}\ }\textbf {\bibinfo {volume} {101}},\ \bibinfo {pages}
  {052003} (\bibinfo {year} {2020})},\ \Eprint
  {http://arxiv.org/abs/1911.03085} {arXiv:1911.03085 [hep-ex]} \BibitemShut
  {NoStop}%
\bibitem [{\citenamefont {Bell}\ \emph {et~al.}(2021)\citenamefont {Bell},
  \citenamefont {Dent}, \citenamefont {Dutta}, \citenamefont {Ghosh},
  \citenamefont {Kumar},\ and\ \citenamefont {Newstead}}]{Bell:2021zkr}%
  \BibitemOpen
  \bibfield  {author} {\bibinfo {author} {\bibfnamefont {N.~F.}\ \bibnamefont
  {Bell}}, \bibinfo {author} {\bibfnamefont {J.~B.}\ \bibnamefont {Dent}},
  \bibinfo {author} {\bibfnamefont {B.}~\bibnamefont {Dutta}}, \bibinfo
  {author} {\bibfnamefont {S.}~\bibnamefont {Ghosh}}, \bibinfo {author}
  {\bibfnamefont {J.}~\bibnamefont {Kumar}}, \ and\ \bibinfo {author}
  {\bibfnamefont {J.~L.}\ \bibnamefont {Newstead}},\ }\href {\doibase
  10.1103/PhysRevD.104.076013} {\bibfield  {journal} {\bibinfo  {journal}
  {Phys. Rev. D}\ }\textbf {\bibinfo {volume} {104}},\ \bibinfo {pages}
  {076013} (\bibinfo {year} {2021})},\ \Eprint
  {http://arxiv.org/abs/2103.05890} {arXiv:2103.05890 [hep-ph]} \BibitemShut
  {NoStop}%
\bibitem [{\citenamefont {Necib}\ and\ \citenamefont
  {Lin}(2021)}]{Necib:2021vxr}%
  \BibitemOpen
  \bibfield  {author} {\bibinfo {author} {\bibfnamefont {L.}~\bibnamefont
  {Necib}}\ and\ \bibinfo {author} {\bibfnamefont {T.}~\bibnamefont {Lin}},\
  }\href@noop {} {\  (\bibinfo {year} {2021})},\ \Eprint
  {http://arxiv.org/abs/2102.02211} {arXiv:2102.02211 [astro-ph.GA]}
  \BibitemShut {NoStop}%
\bibitem [{\citenamefont {{P.A.~Majewski for the MIGDAL
  Collaboration}}(2021)}]{Majewski:2021}%
  \BibitemOpen
  \bibfield  {author} {\bibinfo {author} {\bibnamefont {{P.A.~Majewski for the
  MIGDAL Collaboration}}},\ }\href
  {https://indico.ific.uv.es/event/6178/contributions/15922/} {\enquote
  {\bibinfo {title} {{The MIGDAL experiment : Towards observation and
  measurement of the Migdal effect to help low mass WIMPs searches}},}\ }
  (\bibinfo {year} {2021}),\ \bibinfo {note} {presented at
  TAUP2021}\BibitemShut {NoStop}%
\bibitem [{\citenamefont {Liao}\ \emph {et~al.}(2021)\citenamefont {Liao},
  \citenamefont {Liu},\ and\ \citenamefont {Marfatia}}]{Liao:2021yog}%
  \BibitemOpen
  \bibfield  {author} {\bibinfo {author} {\bibfnamefont {J.}~\bibnamefont
  {Liao}}, \bibinfo {author} {\bibfnamefont {H.}~\bibnamefont {Liu}}, \ and\
  \bibinfo {author} {\bibfnamefont {D.}~\bibnamefont {Marfatia}},\ }\href
  {\doibase 10.1103/PhysRevD.104.015005} {\bibfield  {journal} {\bibinfo
  {journal} {Phys. Rev. D}\ }\textbf {\bibinfo {volume} {104}},\ \bibinfo
  {pages} {015005} (\bibinfo {year} {2021})},\ \Eprint
  {http://arxiv.org/abs/2104.01811} {arXiv:2104.01811 [hep-ph]} \BibitemShut
  {NoStop}%
\bibitem [{\citenamefont {SHIBATA}\ \emph {et~al.}(2011)\citenamefont
  {SHIBATA}, \citenamefont {IWAMOTO}, \citenamefont {NAKAGAWA}, \citenamefont
  {IWAMOTO}, \citenamefont {ICHIHARA}, \citenamefont {KUNIEDA}, \citenamefont
  {CHIBA}, \citenamefont {FURUTAKA}, \citenamefont {OTUKA}, \citenamefont
  {OHSAWA}, \citenamefont {MURATA}, \citenamefont {MATSUNOBU}, \citenamefont
  {ZUKERAN}, \citenamefont {KAMADA},\ and\ \citenamefont {ichi
  KATAKURA}}]{JENDL4.0:2011}%
  \BibitemOpen
  \bibfield  {author} {\bibinfo {author} {\bibfnamefont {K.}~\bibnamefont
  {SHIBATA}}, \bibinfo {author} {\bibfnamefont {O.}~\bibnamefont {IWAMOTO}},
  \bibinfo {author} {\bibfnamefont {T.}~\bibnamefont {NAKAGAWA}}, \bibinfo
  {author} {\bibfnamefont {N.}~\bibnamefont {IWAMOTO}}, \bibinfo {author}
  {\bibfnamefont {A.}~\bibnamefont {ICHIHARA}}, \bibinfo {author}
  {\bibfnamefont {S.}~\bibnamefont {KUNIEDA}}, \bibinfo {author} {\bibfnamefont
  {S.}~\bibnamefont {CHIBA}}, \bibinfo {author} {\bibfnamefont
  {K.}~\bibnamefont {FURUTAKA}}, \bibinfo {author} {\bibfnamefont
  {N.}~\bibnamefont {OTUKA}}, \bibinfo {author} {\bibfnamefont
  {T.}~\bibnamefont {OHSAWA}}, \bibinfo {author} {\bibfnamefont
  {T.}~\bibnamefont {MURATA}}, \bibinfo {author} {\bibfnamefont
  {H.}~\bibnamefont {MATSUNOBU}}, \bibinfo {author} {\bibfnamefont
  {A.}~\bibnamefont {ZUKERAN}}, \bibinfo {author} {\bibfnamefont
  {S.}~\bibnamefont {KAMADA}}, \ and\ \bibinfo {author} {\bibfnamefont
  {J.}~\bibnamefont {ichi KATAKURA}},\ }\href {\doibase
  10.1080/18811248.2011.9711675} {\bibfield  {journal} {\bibinfo  {journal}
  {Journal of Nuclear Science and Technology}\ }\textbf {\bibinfo {volume}
  {48}},\ \bibinfo {pages} {1} (\bibinfo {year} {2011})},\ \Eprint
  {http://arxiv.org/abs/https://doi.org/10.1080/18811248.2011.9711675}
  {https://doi.org/10.1080/18811248.2011.9711675} \BibitemShut {NoStop}%
\bibitem [{\citenamefont {Nakamura}\ \emph {et~al.}(2021)\citenamefont
  {Nakamura}, \citenamefont {Miuchi}, \citenamefont {Kazama}, \citenamefont
  {Shoji}, \citenamefont {Ibe},\ and\ \citenamefont
  {Nakano}}]{Nakamura:2020kex}%
  \BibitemOpen
  \bibfield  {author} {\bibinfo {author} {\bibfnamefont {K.~D.}\ \bibnamefont
  {Nakamura}}, \bibinfo {author} {\bibfnamefont {K.}~\bibnamefont {Miuchi}},
  \bibinfo {author} {\bibfnamefont {S.}~\bibnamefont {Kazama}}, \bibinfo
  {author} {\bibfnamefont {Y.}~\bibnamefont {Shoji}}, \bibinfo {author}
  {\bibfnamefont {M.}~\bibnamefont {Ibe}}, \ and\ \bibinfo {author}
  {\bibfnamefont {W.}~\bibnamefont {Nakano}},\ }\href {\doibase
  10.1093/ptep/ptaa162} {\bibfield  {journal} {\bibinfo  {journal} {PTEP}\
  }\textbf {\bibinfo {volume} {2021}},\ \bibinfo {pages} {013C01} (\bibinfo
  {year} {2021})},\ \Eprint {http://arxiv.org/abs/2009.05939} {arXiv:2009.05939
  [physics.ins-det]} \BibitemShut {NoStop}%
\bibitem [{\citenamefont {Talman}\ and\ \citenamefont
  {Frolov}(2006)}]{PhysRevA.73.032722}%
  \BibitemOpen
  \bibfield  {author} {\bibinfo {author} {\bibfnamefont {J.~D.}\ \bibnamefont
  {Talman}}\ and\ \bibinfo {author} {\bibfnamefont {A.~M.}\ \bibnamefont
  {Frolov}},\ }\href {\doibase 10.1103/PhysRevA.73.032722} {\bibfield
  {journal} {\bibinfo  {journal} {Phys. Rev. A}\ }\textbf {\bibinfo {volume}
  {73}},\ \bibinfo {pages} {032722} (\bibinfo {year} {2006})}\BibitemShut
  {NoStop}%
\bibitem [{\citenamefont {Akerib}\ \emph {et~al.}(2016)\citenamefont {Akerib}
  \emph {et~al.}}]{Akerib:2016mzi}%
  \BibitemOpen
  \bibfield  {author} {\bibinfo {author} {\bibfnamefont {D.}~\bibnamefont
  {Akerib}} \emph {et~al.} (\bibinfo {collaboration} {LUX}),\ }\href@noop {} {\
   (\bibinfo {year} {2016})},\ \Eprint {http://arxiv.org/abs/1608.05381}
  {arXiv:1608.05381 [physics.ins-det]} \BibitemShut {NoStop}%
\bibitem [{\citenamefont {Akerib}\ \emph
  {et~al.}(2017{\natexlab{a}})\citenamefont {Akerib} \emph
  {et~al.}}]{LUX:2016ggv}%
  \BibitemOpen
  \bibfield  {author} {\bibinfo {author} {\bibfnamefont {D.~S.}\ \bibnamefont
  {Akerib}} \emph {et~al.} (\bibinfo {collaboration} {LUX}),\ }\href {\doibase
  10.1103/PhysRevLett.118.021303} {\bibfield  {journal} {\bibinfo  {journal}
  {Phys. Rev. Lett.}\ }\textbf {\bibinfo {volume} {118}},\ \bibinfo {pages}
  {021303} (\bibinfo {year} {2017}{\natexlab{a}})},\ \Eprint
  {http://arxiv.org/abs/1608.07648} {arXiv:1608.07648 [astro-ph.CO]}
  \BibitemShut {NoStop}%
\bibitem [{\citenamefont {Aprile}\ \emph {et~al.}(2018)\citenamefont {Aprile}
  \emph {et~al.}}]{XENON:2018voc}%
  \BibitemOpen
  \bibfield  {author} {\bibinfo {author} {\bibfnamefont {E.}~\bibnamefont
  {Aprile}} \emph {et~al.} (\bibinfo {collaboration} {XENON}),\ }\href
  {\doibase 10.1103/PhysRevLett.121.111302} {\bibfield  {journal} {\bibinfo
  {journal} {Phys. Rev. Lett.}\ }\textbf {\bibinfo {volume} {121}},\ \bibinfo
  {pages} {111302} (\bibinfo {year} {2018})},\ \Eprint
  {http://arxiv.org/abs/1805.12562} {arXiv:1805.12562 [astro-ph.CO]}
  \BibitemShut {NoStop}%
\bibitem [{\citenamefont {Akimov}\ \emph
  {et~al.}(2021{\natexlab{a}})\citenamefont {Akimov} \emph
  {et~al.}}]{COHERENT:2020iec}%
  \BibitemOpen
  \bibfield  {author} {\bibinfo {author} {\bibfnamefont {D.}~\bibnamefont
  {Akimov}} \emph {et~al.} (\bibinfo {collaboration} {COHERENT}),\ }\href
  {\doibase 10.1103/PhysRevLett.126.012002} {\bibfield  {journal} {\bibinfo
  {journal} {Phys. Rev. Lett.}\ }\textbf {\bibinfo {volume} {126}},\ \bibinfo
  {pages} {012002} (\bibinfo {year} {2021}{\natexlab{a}})},\ \Eprint
  {http://arxiv.org/abs/2003.10630} {arXiv:2003.10630 [nucl-ex]} \BibitemShut
  {NoStop}%
\bibitem [{\citenamefont {Agnes}\ \emph {et~al.}(2018)\citenamefont {Agnes}
  \emph {et~al.}}]{DarkSide:2018kuk}%
  \BibitemOpen
  \bibfield  {author} {\bibinfo {author} {\bibfnamefont {P.}~\bibnamefont
  {Agnes}} \emph {et~al.} (\bibinfo {collaboration} {DarkSide}),\ }\href
  {\doibase 10.1103/PhysRevD.98.102006} {\bibfield  {journal} {\bibinfo
  {journal} {Phys. Rev. D}\ }\textbf {\bibinfo {volume} {98}},\ \bibinfo
  {pages} {102006} (\bibinfo {year} {2018})},\ \Eprint
  {http://arxiv.org/abs/1802.07198} {arXiv:1802.07198 [astro-ph.CO]}
  \BibitemShut {NoStop}%
\bibitem [{\citenamefont {Barbeau}\ \emph {et~al.}(2007)\citenamefont
  {Barbeau}, \citenamefont {Collar},\ and\ \citenamefont
  {Whaley}}]{Barbeau:2007qh}%
  \BibitemOpen
  \bibfield  {author} {\bibinfo {author} {\bibfnamefont {P.}~\bibnamefont
  {Barbeau}}, \bibinfo {author} {\bibfnamefont {J.}~\bibnamefont {Collar}}, \
  and\ \bibinfo {author} {\bibfnamefont {P.}~\bibnamefont {Whaley}},\ }\href
  {\doibase 10.1016/j.nima.2007.01.169} {\bibfield  {journal} {\bibinfo
  {journal} {Nucl. Instrum. Meth. A}\ }\textbf {\bibinfo {volume} {574}},\
  \bibinfo {pages} {385} (\bibinfo {year} {2007})},\ \Eprint
  {http://arxiv.org/abs/nucl-ex/0701011} {arXiv:nucl-ex/0701011} \BibitemShut
  {NoStop}%
\bibitem [{\citenamefont {Joshi}\ \emph {et~al.}(2014)\citenamefont {Joshi},
  \citenamefont {Sangiorgio}, \citenamefont {Mozin}, \citenamefont {Norman},
  \citenamefont {Sorensen}, \citenamefont {Foxe}, \citenamefont {Bench},\ and\
  \citenamefont {Bernstein}}]{Joshi:2014oda}%
  \BibitemOpen
  \bibfield  {author} {\bibinfo {author} {\bibfnamefont {T.}~\bibnamefont
  {Joshi}}, \bibinfo {author} {\bibfnamefont {S.}~\bibnamefont {Sangiorgio}},
  \bibinfo {author} {\bibfnamefont {V.}~\bibnamefont {Mozin}}, \bibinfo
  {author} {\bibfnamefont {E.}~\bibnamefont {Norman}}, \bibinfo {author}
  {\bibfnamefont {P.}~\bibnamefont {Sorensen}}, \bibinfo {author}
  {\bibfnamefont {M.}~\bibnamefont {Foxe}}, \bibinfo {author} {\bibfnamefont
  {G.}~\bibnamefont {Bench}}, \ and\ \bibinfo {author} {\bibfnamefont
  {A.}~\bibnamefont {Bernstein}},\ }\href {\doibase 10.1016/j.nimb.2014.04.008}
  {\bibfield  {journal} {\bibinfo  {journal} {Nucl. Instrum. Meth. B}\ }\textbf
  {\bibinfo {volume} {333}},\ \bibinfo {pages} {6} (\bibinfo {year} {2014})},\
  \Eprint {http://arxiv.org/abs/1403.1285} {arXiv:1403.1285 [physics.ins-det]}
  \BibitemShut {NoStop}%
\bibitem [{\citenamefont {Aprile}\ \emph {et~al.}(2020)\citenamefont {Aprile}
  \emph {et~al.}}]{XENON:2020rca}%
  \BibitemOpen
  \bibfield  {author} {\bibinfo {author} {\bibfnamefont {E.}~\bibnamefont
  {Aprile}} \emph {et~al.} (\bibinfo {collaboration} {XENON}),\ }\href
  {\doibase 10.1103/PhysRevD.102.072004} {\bibfield  {journal} {\bibinfo
  {journal} {Phys. Rev. D}\ }\textbf {\bibinfo {volume} {102}},\ \bibinfo
  {pages} {072004} (\bibinfo {year} {2020})},\ \Eprint
  {http://arxiv.org/abs/2006.09721} {arXiv:2006.09721 [hep-ex]} \BibitemShut
  {NoStop}%
\bibitem [{\citenamefont {Helm}(1956)}]{Helm:1956zz}%
  \BibitemOpen
  \bibfield  {author} {\bibinfo {author} {\bibfnamefont {R.~H.}\ \bibnamefont
  {Helm}},\ }\href {\doibase 10.1103/PhysRev.104.1466} {\bibfield  {journal}
  {\bibinfo  {journal} {Phys. Rev.}\ }\textbf {\bibinfo {volume} {104}},\
  \bibinfo {pages} {1466} (\bibinfo {year} {1956})}\BibitemShut {NoStop}%
\bibitem [{\citenamefont {Akimov}\ \emph
  {et~al.}(2021{\natexlab{b}})\citenamefont {Akimov} \emph
  {et~al.}}]{Akimov:2021geg}%
  \BibitemOpen
  \bibfield  {author} {\bibinfo {author} {\bibfnamefont {D.}~\bibnamefont
  {Akimov}} \emph {et~al.},\ }\href@noop {} {\  (\bibinfo {year}
  {2021}{\natexlab{b}})},\ \Eprint {http://arxiv.org/abs/2109.11049}
  {arXiv:2109.11049 [hep-ex]} \BibitemShut {NoStop}%
\bibitem [{\citenamefont {Anselmann}\ \emph {et~al.}(1995)\citenamefont
  {Anselmann}, \citenamefont {Fockenbrock}, \citenamefont {Hampel},
  \citenamefont {Heusser}, \citenamefont {Kiko}, \citenamefont {Kirsten},
  \citenamefont {Laubenstein}, \citenamefont {Pernicka}, \citenamefont
  {Pezzoni}, \citenamefont {Rönn}, \citenamefont {Sann}, \citenamefont
  {Spielker}, \citenamefont {Wink}, \citenamefont {Wójcik}, \citenamefont
  {Ammon}, \citenamefont {Ebert}, \citenamefont {Fritsch}, \citenamefont
  {Heidt}, \citenamefont {Henrich}, \citenamefont {Schlosser}, \citenamefont
  {Stieglitz}, \citenamefont {Weirich}, \citenamefont {Balata}, \citenamefont
  {Lalla}, \citenamefont {Bellotti}, \citenamefont {Cattadori}, \citenamefont
  {Cremonesi}, \citenamefont {Ferrari}, \citenamefont {Fiorini}, \citenamefont
  {Zanotti}, \citenamefont {Altmann}, \citenamefont {Feilitzsch}, \citenamefont
  {Mößbauer}, \citenamefont {Schanda}, \citenamefont {Berthomieu},
  \citenamefont {Schatzman}, \citenamefont {Carmi}, \citenamefont {Dostrovsky},
  \citenamefont {Bacci}, \citenamefont {Belli}, \citenamefont {Bernabei},
  \citenamefont {d'Angelo}, \citenamefont {Paoluzi}, \citenamefont
  {Bevilacqua}, \citenamefont {Charbit}, \citenamefont {Cribier}, \citenamefont
  {Dupont}, \citenamefont {Gosset}, \citenamefont {Rich}, \citenamefont
  {Spiro}, \citenamefont {Stolarczyk}, \citenamefont {Tao}, \citenamefont
  {Vignaud}, \citenamefont {Boger}, \citenamefont {Hahn}, \citenamefont
  {Hartmann}, \citenamefont {Rowley}, \citenamefont {Stoenner},\ and\
  \citenamefont {Weneser}}]{ANSELMANN1995440}%
  \BibitemOpen
  \bibfield  {author} {\bibinfo {author} {\bibfnamefont {P.}~\bibnamefont
  {Anselmann}}, \bibinfo {author} {\bibfnamefont {R.}~\bibnamefont
  {Fockenbrock}}, \bibinfo {author} {\bibfnamefont {W.}~\bibnamefont {Hampel}},
  \bibinfo {author} {\bibfnamefont {G.}~\bibnamefont {Heusser}}, \bibinfo
  {author} {\bibfnamefont {J.}~\bibnamefont {Kiko}}, \bibinfo {author}
  {\bibfnamefont {T.}~\bibnamefont {Kirsten}}, \bibinfo {author} {\bibfnamefont
  {M.}~\bibnamefont {Laubenstein}}, \bibinfo {author} {\bibfnamefont
  {E.}~\bibnamefont {Pernicka}}, \bibinfo {author} {\bibfnamefont
  {S.}~\bibnamefont {Pezzoni}}, \bibinfo {author} {\bibfnamefont
  {U.}~\bibnamefont {Rönn}}, \bibinfo {author} {\bibfnamefont
  {M.}~\bibnamefont {Sann}}, \bibinfo {author} {\bibfnamefont {F.}~\bibnamefont
  {Spielker}}, \bibinfo {author} {\bibfnamefont {R.}~\bibnamefont {Wink}},
  \bibinfo {author} {\bibfnamefont {M.}~\bibnamefont {Wójcik}}, \bibinfo
  {author} {\bibfnamefont {R.}~\bibnamefont {Ammon}}, \bibinfo {author}
  {\bibfnamefont {K.}~\bibnamefont {Ebert}}, \bibinfo {author} {\bibfnamefont
  {T.}~\bibnamefont {Fritsch}}, \bibinfo {author} {\bibfnamefont
  {D.}~\bibnamefont {Heidt}}, \bibinfo {author} {\bibfnamefont
  {E.}~\bibnamefont {Henrich}}, \bibinfo {author} {\bibfnamefont
  {C.}~\bibnamefont {Schlosser}}, \bibinfo {author} {\bibfnamefont
  {L.}~\bibnamefont {Stieglitz}}, \bibinfo {author} {\bibfnamefont
  {F.}~\bibnamefont {Weirich}}, \bibinfo {author} {\bibfnamefont
  {M.}~\bibnamefont {Balata}}, \bibinfo {author} {\bibfnamefont
  {H.}~\bibnamefont {Lalla}}, \bibinfo {author} {\bibfnamefont
  {E.}~\bibnamefont {Bellotti}}, \bibinfo {author} {\bibfnamefont
  {C.}~\bibnamefont {Cattadori}}, \bibinfo {author} {\bibfnamefont
  {O.}~\bibnamefont {Cremonesi}}, \bibinfo {author} {\bibfnamefont
  {N.}~\bibnamefont {Ferrari}}, \bibinfo {author} {\bibfnamefont
  {E.}~\bibnamefont {Fiorini}}, \bibinfo {author} {\bibfnamefont
  {L.}~\bibnamefont {Zanotti}}, \bibinfo {author} {\bibfnamefont
  {M.}~\bibnamefont {Altmann}}, \bibinfo {author} {\bibfnamefont
  {F.}~\bibnamefont {Feilitzsch}}, \bibinfo {author} {\bibfnamefont
  {R.}~\bibnamefont {Mößbauer}}, \bibinfo {author} {\bibfnamefont
  {U.}~\bibnamefont {Schanda}}, \bibinfo {author} {\bibfnamefont
  {G.}~\bibnamefont {Berthomieu}}, \bibinfo {author} {\bibfnamefont
  {E.}~\bibnamefont {Schatzman}}, \bibinfo {author} {\bibfnamefont
  {I.}~\bibnamefont {Carmi}}, \bibinfo {author} {\bibfnamefont
  {I.}~\bibnamefont {Dostrovsky}}, \bibinfo {author} {\bibfnamefont
  {C.}~\bibnamefont {Bacci}}, \bibinfo {author} {\bibfnamefont
  {P.}~\bibnamefont {Belli}}, \bibinfo {author} {\bibfnamefont
  {R.}~\bibnamefont {Bernabei}}, \bibinfo {author} {\bibfnamefont
  {S.}~\bibnamefont {d'Angelo}}, \bibinfo {author} {\bibfnamefont
  {L.}~\bibnamefont {Paoluzi}}, \bibinfo {author} {\bibfnamefont
  {A.}~\bibnamefont {Bevilacqua}}, \bibinfo {author} {\bibfnamefont
  {S.}~\bibnamefont {Charbit}}, \bibinfo {author} {\bibfnamefont
  {M.}~\bibnamefont {Cribier}}, \bibinfo {author} {\bibfnamefont
  {G.}~\bibnamefont {Dupont}}, \bibinfo {author} {\bibfnamefont
  {L.}~\bibnamefont {Gosset}}, \bibinfo {author} {\bibfnamefont
  {J.}~\bibnamefont {Rich}}, \bibinfo {author} {\bibfnamefont {M.}~\bibnamefont
  {Spiro}}, \bibinfo {author} {\bibfnamefont {T.}~\bibnamefont {Stolarczyk}},
  \bibinfo {author} {\bibfnamefont {C.}~\bibnamefont {Tao}}, \bibinfo {author}
  {\bibfnamefont {D.}~\bibnamefont {Vignaud}}, \bibinfo {author} {\bibfnamefont
  {J.}~\bibnamefont {Boger}}, \bibinfo {author} {\bibfnamefont
  {R.}~\bibnamefont {Hahn}}, \bibinfo {author} {\bibfnamefont {F.}~\bibnamefont
  {Hartmann}}, \bibinfo {author} {\bibfnamefont {J.}~\bibnamefont {Rowley}},
  \bibinfo {author} {\bibfnamefont {R.}~\bibnamefont {Stoenner}}, \ and\
  \bibinfo {author} {\bibfnamefont {J.}~\bibnamefont {Weneser}},\ }\href
  {\doibase https://doi.org/10.1016/0370-2693(94)01586-2} {\bibfield  {journal}
  {\bibinfo  {journal} {Physics Letters B}\ }\textbf {\bibinfo {volume}
  {342}},\ \bibinfo {pages} {440} (\bibinfo {year} {1995})}\BibitemShut
  {NoStop}%
\bibitem [{\citenamefont {Cribier}\ \emph {et~al.}(1988)\citenamefont
  {Cribier}, \citenamefont {Pichard}, \citenamefont {Rich}, \citenamefont
  {Spiro}, \citenamefont {Vignaud}, \citenamefont {Besson}, \citenamefont
  {Bevilacqua}, \citenamefont {Caperan}, \citenamefont {Dupont}, \citenamefont
  {Sire}, \citenamefont {Gorry}, \citenamefont {Hampel},\ and\ \citenamefont
  {Kirsten}}]{CRIBIER1988574}%
  \BibitemOpen
  \bibfield  {author} {\bibinfo {author} {\bibfnamefont {M.}~\bibnamefont
  {Cribier}}, \bibinfo {author} {\bibfnamefont {B.}~\bibnamefont {Pichard}},
  \bibinfo {author} {\bibfnamefont {J.}~\bibnamefont {Rich}}, \bibinfo {author}
  {\bibfnamefont {M.}~\bibnamefont {Spiro}}, \bibinfo {author} {\bibfnamefont
  {D.}~\bibnamefont {Vignaud}}, \bibinfo {author} {\bibfnamefont
  {A.}~\bibnamefont {Besson}}, \bibinfo {author} {\bibfnamefont
  {A.}~\bibnamefont {Bevilacqua}}, \bibinfo {author} {\bibfnamefont
  {F.}~\bibnamefont {Caperan}}, \bibinfo {author} {\bibfnamefont
  {G.}~\bibnamefont {Dupont}}, \bibinfo {author} {\bibfnamefont
  {P.}~\bibnamefont {Sire}}, \bibinfo {author} {\bibfnamefont {J.}~\bibnamefont
  {Gorry}}, \bibinfo {author} {\bibfnamefont {W.}~\bibnamefont {Hampel}}, \
  and\ \bibinfo {author} {\bibfnamefont {T.}~\bibnamefont {Kirsten}},\ }\href
  {\doibase https://doi.org/10.1016/S0168-9002(98)90030-4} {\bibfield
  {journal} {\bibinfo  {journal} {Nuclear Instruments and Methods in Physics
  Research Section A: Accelerators, Spectrometers, Detectors and Associated
  Equipment}\ }\textbf {\bibinfo {volume} {265}},\ \bibinfo {pages} {574}
  (\bibinfo {year} {1988})}\BibitemShut {NoStop}%
\bibitem [{\citenamefont {Akerib}\ \emph {et~al.}(2013)\citenamefont {Akerib}
  \emph {et~al.}}]{LUX:2012kmp}%
  \BibitemOpen
  \bibfield  {author} {\bibinfo {author} {\bibfnamefont {D.~S.}\ \bibnamefont
  {Akerib}} \emph {et~al.} (\bibinfo {collaboration} {LUX}),\ }\href {\doibase
  10.1016/j.nima.2012.11.135} {\bibfield  {journal} {\bibinfo  {journal} {Nucl.
  Instrum. Meth. A}\ }\textbf {\bibinfo {volume} {704}},\ \bibinfo {pages}
  {111} (\bibinfo {year} {2013})},\ \Eprint {http://arxiv.org/abs/1211.3788}
  {arXiv:1211.3788 [physics.ins-det]} \BibitemShut {NoStop}%
\bibitem [{\citenamefont {Akerib}\ \emph {et~al.}(2020)\citenamefont {Akerib}
  \emph {et~al.}}]{LUX:2020car}%
  \BibitemOpen
  \bibfield  {author} {\bibinfo {author} {\bibfnamefont {D.~S.}\ \bibnamefont
  {Akerib}} \emph {et~al.} (\bibinfo {collaboration} {LUX}),\ }\href {\doibase
  10.1103/PhysRevD.102.112002} {\bibfield  {journal} {\bibinfo  {journal}
  {Phys. Rev. D}\ }\textbf {\bibinfo {volume} {102}},\ \bibinfo {pages}
  {112002} (\bibinfo {year} {2020})},\ \Eprint
  {http://arxiv.org/abs/2004.06304} {arXiv:2004.06304 [physics.ins-det]}
  \BibitemShut {NoStop}%
\bibitem [{\citenamefont {Szydagis}\ \emph {et~al.}(2011)\citenamefont
  {Szydagis}, \citenamefont {Barry}, \citenamefont {Kazkaz}, \citenamefont
  {Mock}, \citenamefont {Stolp}, \citenamefont {Sweany}, \citenamefont
  {Tripathi}, \citenamefont {Uvarov}, \citenamefont {Walsh},\ and\
  \citenamefont {Woods}}]{Szydagis_2011}%
  \BibitemOpen
  \bibfield  {author} {\bibinfo {author} {\bibfnamefont {M.}~\bibnamefont
  {Szydagis}}, \bibinfo {author} {\bibfnamefont {N.}~\bibnamefont {Barry}},
  \bibinfo {author} {\bibfnamefont {K.}~\bibnamefont {Kazkaz}}, \bibinfo
  {author} {\bibfnamefont {J.}~\bibnamefont {Mock}}, \bibinfo {author}
  {\bibfnamefont {D.}~\bibnamefont {Stolp}}, \bibinfo {author} {\bibfnamefont
  {M.}~\bibnamefont {Sweany}}, \bibinfo {author} {\bibfnamefont
  {M.}~\bibnamefont {Tripathi}}, \bibinfo {author} {\bibfnamefont
  {S.}~\bibnamefont {Uvarov}}, \bibinfo {author} {\bibfnamefont
  {N.}~\bibnamefont {Walsh}}, \ and\ \bibinfo {author} {\bibfnamefont
  {M.}~\bibnamefont {Woods}},\ }\href {\doibase 10.1088/1748-0221/6/10/p10002}
  {\bibfield  {journal} {\bibinfo  {journal} {Journal of Instrumentation}\
  }\textbf {\bibinfo {volume} {6}},\ \bibinfo {pages} {P10002} (\bibinfo {year}
  {2011})}\BibitemShut {NoStop}%
\bibitem [{\citenamefont {Szydagis}\ \emph {et~al.}(2021)\citenamefont
  {Szydagis}, \citenamefont {Balajthy}, \citenamefont {Block}, \citenamefont
  {Brodsky}, \citenamefont {Cutter}, \citenamefont {Farrell}, \citenamefont
  {Huang}, \citenamefont {Kozlova}, \citenamefont {Lenardo}, \citenamefont
  {Manalaysay}, \citenamefont {McKinsey}, \citenamefont {Mooney}, \citenamefont
  {Mueller}, \citenamefont {Ni}, \citenamefont {Rischbieter}, \citenamefont
  {Tripathi}, \citenamefont {Tunnell}, \citenamefont {Velan},\ and\
  \citenamefont {Zhao}}]{szydagis_m_2021_5676553}%
  \BibitemOpen
  \bibfield  {author} {\bibinfo {author} {\bibfnamefont {M.}~\bibnamefont
  {Szydagis}}, \bibinfo {author} {\bibfnamefont {J.}~\bibnamefont {Balajthy}},
  \bibinfo {author} {\bibfnamefont {G.}~\bibnamefont {Block}}, \bibinfo
  {author} {\bibfnamefont {J.}~\bibnamefont {Brodsky}}, \bibinfo {author}
  {\bibfnamefont {J.}~\bibnamefont {Cutter}}, \bibinfo {author} {\bibfnamefont
  {S.}~\bibnamefont {Farrell}}, \bibinfo {author} {\bibfnamefont
  {J.}~\bibnamefont {Huang}}, \bibinfo {author} {\bibfnamefont
  {E.}~\bibnamefont {Kozlova}}, \bibinfo {author} {\bibfnamefont
  {B.}~\bibnamefont {Lenardo}}, \bibinfo {author} {\bibfnamefont
  {A.}~\bibnamefont {Manalaysay}}, \bibinfo {author} {\bibfnamefont
  {D.}~\bibnamefont {McKinsey}}, \bibinfo {author} {\bibfnamefont
  {M.}~\bibnamefont {Mooney}}, \bibinfo {author} {\bibfnamefont
  {J.}~\bibnamefont {Mueller}}, \bibinfo {author} {\bibfnamefont
  {K.}~\bibnamefont {Ni}}, \bibinfo {author} {\bibfnamefont {G.}~\bibnamefont
  {Rischbieter}}, \bibinfo {author} {\bibfnamefont {M.}~\bibnamefont
  {Tripathi}}, \bibinfo {author} {\bibfnamefont {C.}~\bibnamefont {Tunnell}},
  \bibinfo {author} {\bibfnamefont {V.}~\bibnamefont {Velan}}, \ and\ \bibinfo
  {author} {\bibfnamefont {Z.}~\bibnamefont {Zhao}},\ }\href {\doibase
  10.5281/zenodo.5676553} {\enquote {\bibinfo {title} {Noble element simulation
  technique},}\ } (\bibinfo {year} {2021})\BibitemShut {NoStop}%
\bibitem [{\citenamefont {Agency}(2007)}]{agency_database_2007}%
  \BibitemOpen
  \bibfield  {author} {\bibinfo {author} {\bibfnamefont {I.~A.~E.}\
  \bibnamefont {Agency}},\ }\href
  {https://books.google.com.au/books?id=3h9RAAAAMAAJ} {\emph {\bibinfo {title}
  {Database of {Prompt} {Gamma} {Rays} from {Slow} {Neutron} {Capture} for
  {Elemental} {Analysis}}}},\ {STI}/{PUB}\ (\bibinfo  {publisher}
  {International Atomic Energy Agency},\ \bibinfo {year} {2007})\BibitemShut
  {NoStop}%
\bibitem [{\citenamefont {Mougeot}(2015)}]{PhysRevC.91.055504}%
  \BibitemOpen
  \bibfield  {author} {\bibinfo {author} {\bibfnamefont {X.}~\bibnamefont
  {Mougeot}},\ }\href {\doibase 10.1103/PhysRevC.91.055504} {\bibfield
  {journal} {\bibinfo  {journal} {Phys. Rev. C}\ }\textbf {\bibinfo {volume}
  {91}},\ \bibinfo {pages} {055504} (\bibinfo {year} {2015})}\BibitemShut
  {NoStop}%
\bibitem [{\citenamefont {Akerib}\ \emph
  {et~al.}(2017{\natexlab{b}})\citenamefont {Akerib} \emph
  {et~al.}}]{LUX:2017ojt}%
  \BibitemOpen
  \bibfield  {author} {\bibinfo {author} {\bibfnamefont {D.~S.}\ \bibnamefont
  {Akerib}} \emph {et~al.} (\bibinfo {collaboration} {LUX}),\ }\href {\doibase
  10.1103/PhysRevD.96.112011} {\bibfield  {journal} {\bibinfo  {journal} {Phys.
  Rev. D}\ }\textbf {\bibinfo {volume} {96}},\ \bibinfo {pages} {112011}
  (\bibinfo {year} {2017}{\natexlab{b}})},\ \Eprint
  {http://arxiv.org/abs/1709.00800} {arXiv:1709.00800 [physics.ins-det]}
  \BibitemShut {NoStop}%
\end{thebibliography}%

\end{document}